\documentstyle[aps,preprint,eqsecnum,epsfig]{revtex}
\begin{document}
\draft 
\tighten
\preprint{UFIFT-HEP-98-12}
\date{\today}

\title{Studies of the motion and decay of \\
axion walls bounded by strings.}
\author{S. Chang$^{1}$, C. Hagmann$^{2}$ and P. Sikivie$^{1}$}
\address{
$^{1}$Department of Physics, University of Florida, Gainesville, FL 32611}
\address{
$^{2}$Lawrence Livermore National Laboratory, Livermore, CA 94550}

\maketitle

\begin{abstract}
We discuss the appearance at the QCD phase transition, and the
subsequent decay, of axion walls bounded by strings in $N=1$
axion models.  We argue on intuitive grounds that the main 
decay mechanism is into barely relativistic axions.  We present 
numerical simulations of the decay process.  In these simulations, 
the decay happens immediately, in a time scale of order the 
light travel time, and the average energy of the radiated axions
is $\langle \omega_a \rangle \simeq 7 m_a$ for $v_a/m_a \simeq 500$.
$\langle \omega_a \rangle$ is found to increase approximately 
linearly with $\ln(v_a/m_a)$.  Extrapolation of this behaviour yields 
$\langle \omega_a \rangle \sim 60~m_a$ in axion models of interest. 
We find that the contribution to the cosmological energy density 
of axions from wall decay is of the same order of magnitude as that 
from vacuum realignment, with however large uncertainties.  The 
velocity dispersion of axions from wall decay is found to be 
larger, by a factor $10^3$ or so, than that of axions from vacuum 
realignment and string decay.  We discuss the implications of 
this for the formation and evolution of axion miniclusters and for 
the direct detection of axion dark matter on Earth.  Finally we 
discuss the cosmology of axion models with $N>1$ in which the 
domain wall problem is solved by introducing a small U$_{PQ}$(1) 
breaking interaction.  We find that in this case the walls decay 
into gravitational waves.   
\end{abstract}


\narrowtext

\section{Introduction}
\label{sec:in}

The axion \cite{PQWW,reviews} was postulated approximately twenty years
ago to explain why the strong interactions conserve the discrete symmetries 
P and CP in spite of the fact that the Standard Model of particle interactions 
as a whole violates those symmetries.  It is the quasi-Nambu-Goldstone
boson associated with the spontaneous breaking of a $U_{PQ}(1)$ symmetry
which Peccei and Quinn postulated.  At zero temperature the axion mass is
given by:
\begin{equation}
m_a\simeq 6 \cdot 10^{-6} \mbox{eV} \cdot N \cdot \left(\frac{10^{12}
\mbox{GeV}} {v_a} \right) \label{1.1}
\end{equation}
where $v_a$ is the magnitude of the vacuum expectation value that breaks
$U_{PQ}(1)$ and $N$ is a strictly positive integer that describes the 
color anomaly of $U_{PQ}(1)$.  Axion models have $N$ degenerate vacua
\cite{degen,reviews}.  Searches for the axion in high energy and nuclear
physics experiments have only produced negative results.  By combining
the constraints from these experiments with those from astrophysics 
\cite{reviews,astro}, one obtains the following bound: 
$m_a \lesssim 10^{-2}$ eV.  

The axion owes its mass to non-perturbative QCD effects.  In the very early
universe, at temperatures high compared to the QCD scale, these effects are 
suppressed \cite{highT} and the axion mass is negligible. The axion 
mass turns on when the temperature approaches the QCD scale and increases 
till it reaches the value given in Eq.(1.1) which is valid below the QCD 
scale.  There is a critical time $t_1$, defined by $m_a(t_1) t_1 = 1$, when 
the axion mass effectively turns on \cite{vacmis}. The corresponding 
temperature $T_1 \simeq 1$ GeV.

The implications of the existence of an axion for the history of the 
early universe may be briefly described as follows.  At a temperature of 
order $v_a$, a phase transition occurs in which the $U_{PQ}(1)$ symmetry 
becomes spontaneously broken.  This is called the PQ phase transition.  At 
that time axion strings appear as topological defects.  One must distinguish 
two cases: 1) inflation occurs with reheat temperature higher than the PQ 
transition temperature (equivalently, for the purposes of this paper, 
inflation does not occur at all) or 2) inflation occurs with reheat 
temperature less than the PQ transition temperature.
  
In case 2) the axion field gets homogenized by inflation and the axion 
strings are `blown away'.  When the axion mass turns on at $t_1$, the
axion field starts to oscillate.  The amplitude of this oscillation
is determined by how far from zero the axion field is when the axion 
mass turns on.  The axion field oscillations do not dissipate into 
other forms of energy and hence contribute to the cosmological 
energy density today \cite{vacmis}.  Such a contribution is called of 
``vacuum realignment".  Note that the vacuum realignment contribution 
may be accidentally suppressed in case 2) because the axion field, which 
has been homogenized by inflation, may happen to lie close to zero.

In case 1) the axion strings radiate axions \cite{Davis,Harari} from the 
time of the PQ transition till $t_1$ when the axion mass turns on.   At
$t_1$ each string becomes the boundary of $N$ domain walls.  If $N=1$, 
the network of walls bounded by strings is unstable \cite{Vil,Paris} and
decays away.  All of the present paper except section VI is concerned 
with this process, how many axions are produced in the decay, what is the
velocity dispersion of these axions and what are the implications.

If $N>1$ there is a domain wall problem \cite{degen} because axion domain 
walls end up dominating the energy density, resulting in a universe very
different from the one observed today.  There is a way to avoid the domain 
wall problem by introducing an interaction which slightly lowers one of 
the $N$ vacua with respect to the others.  In that case, the lowest vacuum 
takes over after some time and the domain walls disappear.  There is  
little room in parameter space for that to happen but it is a logical 
possibility.  Section VI discusses this in detail.

In case 1) there are three contributions to the axion cosmological 
energy density.  One contribution \cite{Davis,Harari,stringA,Hagmann}
is from axions that were radiated by axion strings before $t_1$; let us 
call it the string decay contribution.  A second contribution is from 
axions that were produced in the decay of walls bounded by strings after 
$t_1$ \cite{Hagmann,Lyth,Nagasawa}; call it the contribution from wall 
decay.  A third contribution is from vacuum realignment \cite{vacmis}.  
To convince oneself that there is a vacuum realignment contribution distinct 
from the other two, consider a region of the universe which happens to be 
free of strings and domain walls.  In such a region the axion field is 
generally different from zero, even though no strings or walls are present.  
After time $t_1$, the axion field oscillates implying a contribution to the
energy density which is neither from string decay nor wall decay.  Since 
the axion field oscillations caused by vacuum realignment, string decay 
and wall decay are mutually incoherent, the three contributions to the 
energy density should simply be added to each other \cite{Hagmann}.

That axions walls bounded by strings decay predominantly into barely 
relativistic axions was suggested in ref.\cite{Hagmann}, where the size of 
the wall contribution to the axion energy density was estimated to be of 
the same order of magnitude as that from string decay and from vacuum 
realignment.  This is consistent with what we find here.  D.~Lyth \cite{Lyth}
also discussed the wall decay contribution and emphasized the uncertainties 
affecting it.  M.~Nagasawa and M.~Kawasaki \cite{Nagasawa} performed
computer simulations of the decay of walls bounded by string and obtained 
$\langle \omega_a \rangle /m_a \simeq 3$ for the average energy of the 
radiated axions.  Our simulations, presented in section IV, are similar to 
those of Nagasawa and Kawasaki but they are done on larger lattices and
for a wider variety of initial conditions.  We obtain 
$\langle \omega_a \rangle/ m_a \simeq 7$ for $v_a/m_a \simeq 500$.  We
attribute the difference between our result and that of Nagasawa and Kawasaki 
to the fact that we give angular momentum to the collapsing walls, whereas 
they did not.  

It should be emphasized that the lattice sizes which are amenable to
present day computers are at any rate small compared to what one 
would ideally wish.  Indeed the axion string core has size of order
$\frac{1}{\sqrt{\lambda} v_a}$ where $\lambda$ is a coupling constant
(see section \ref{wbs}) whereas the axion domain wall thickness is of    
order $m_a^{-1}$.  The lattice constant must be smaller than
$\frac{1}{\sqrt{\lambda}v_a}$ for the lattice to resolve the string 
core.  On the other hand the lattice size must be larger than
$m_a^{-1}$ to contain at least one wall. Hence the lattice size in units
of the lattice constant must be of order
$10\frac{\sqrt{\lambda}v_a}{m_a} \times 10\frac{\sqrt{\lambda}v_a}{m_a}$ 
or larger if the simulations are done in 2 dimensions (see section \ref{sos}).  
Present day computers allow lattice sizes of order $4000\times 4000$, i.e. 
$\frac{\sqrt{\lambda}v_a}{m_a}\sim 100$.  In axion models of interest
$\frac{\sqrt{\lambda}v_a}{m_a} \sim \frac{10^{12}{\rm GeV}}{10^{-5}{\rm eV}}
=10^{26}$.   The computer simulations inform us about the situation of 
interest only insofar it may be assumed that the motion and decay of walls 
bounded by strings do not vary dramatically from the case where 
$\frac{v_a}{m_a}$ is large to the case where $\frac{v_a}{m_a}$ is huge. 
To address this issue, we study the dependence of 
$\langle \omega_a \rangle / m_a$ upon the $\frac{\sqrt{\lambda}v_a}{m_a}$
and find that it increases approximately as the logarithm of that 
quantity over the range accessible to the simulations.  This behaviour 
is understood in terms of the process by which the walls bounded by 
string decay.  If it persists all the way up to 
$\frac{\sqrt{\lambda}v_a}{m_a} = 10^{26}$, then 
$\langle \omega_a \rangle / m_a$ is of order 60 in models of interest.

There has been disagreement in the literature about the size of the 
string contribution.  Many authors \cite{Davis,stringA} believe it is about 
a factor 100 larger than the vacuum realignment contribution.  We ourselves 
\cite{Harari,Hagmann} have argued that the string and vacuum realignment 
contributions are of the same order of magnitude.  The two different 
estimates are presented in subsection IIIB.  With regard to the wall 
contribution, we argue that it is again of the same order of magnitude 
as that from vacuum realignment, with however large uncertainties.  
Section III presents a unified treatment of all contributions.  The 
resulting axion cosmological energy density is obtained in subsection 
IIID.  The requirement that it does not overclose the universe, implies 
$m_a \gtrsim 10^{-6}$ eV.  By combining this bound with the one from stellar 
evolution mentioned earlier, the axion mass is constrained to lie in the 
window: $10^{-2}$eV $\gtrsim m_a \gtrsim 10^{-6}$ eV.

We find that the velocity dispersion of the axions from wall decay is much 
larger, by a factor of $10^3$ or so, than the velocity dispersion of the 
axions from string decay and vacuum realignment.  This has interesting 
implications for the formation and evolution of axion mini-clusters 
\cite{Hogan,Kolb} as discussed in section V.  We find that the axions 
from wall decay bind only very loosely to axion mini-clusters and that 
they get readily stripped off when the mini-cluster falls into a 
galactic halo.  This ensures the existence of an unclustered component 
of galactic axion dark matter.  The existence of such a component is 
important for searches of axion dark matter on Earth \cite{experim} 
because it guarantees that the signal is on at all times.

Another interesting consequence of the larger velocity dispersion of the 
axions from wall decay is that it may conceivably be measured.  If a signal 
is found in the cavity detector of dark matter axions \cite{experim}, the 
energy spectrum of these axions can be measured with great resolution. It 
has been pointed out that there are peaks in the spectrum \cite{peaks} 
because late infall of cold dark matter onto our galaxy produces distinct 
flows, each one with a characteristic local velocity vector.  These peaks 
are broadened by the primordial velocity dispersion.  The minimum time 
required to measure the primordial velocity dispersion of axions from vacuum 
realignment and string decay is too long ($\sim 10$~years) to be practical, 
but the minimum time required to measure that of axions from wall decay is 
much less.  This is discussed in subsection IIIE.  

This paper is organized as follows. In section II, we present 
a toy model which captures all the physics we are interested in.  We 
discuss the evolution of axion strings in the early universe and 
the formation of walls bounded by strings at the onset of the QCD phase 
transition.  We give an estimate of the density of walls at that time. 
In section III, we discuss the decay of walls bounded by string 
and the axion population that results therefrom.  We also discuss 
the axions from vacuum realignment and from string decay and estimate 
the total cosmological energy density from these mechanisms.  In section 
IV, we present our numerical simulations of the motion and decay of walls 
bounded by strings.  In section V, we discuss the effect of axions from 
wall decay upon the formation and evolution of axion miniclusters.  In 
section VI, we discuss axion models in which $N>1$ and in which the axion 
domain wall problem is solved by slightly lowering one of the $N$ vacua 
with respect to the others.  In section VII, we summarize our conclusions.

\section{The network of walls bounded by strings}
\label{wbs}

The bulk of this paper is concerned with the cosmological properties of 
$N=1$ axion models. $N=1$ means that the axion model has a unique vacuum. 
The following toy model incorporates the qualitative features of interest 
to us:
\begin{eqnarray}
{\cal L} &=&\frac{1}{2} \partial_{\mu} \phi^\dagger \partial^{\mu} \phi
-\frac{\lambda}{4}(\phi^\dagger\phi-v_a^2)^2 + \eta v_a\phi_1 \nonumber\\
&=&\frac{1}{2} \partial_{\mu} \phi_1 \partial^{\mu} \phi_1 +
\frac{1}{2}\partial_{\mu} \phi_2 \partial^{\mu} \phi_2 -\frac{\lambda}{4}
(\phi_1^2+\phi_2^2 -v_a^2)^2 + \eta v_a\phi_1
\label{vi1}
\end{eqnarray}
where $\phi=\phi_1+ i\phi_2$ is a complex scalar field. When $\eta=0$, 
this model has a $U(1)$ global symmetry under which 
$\phi(x) \rightarrow e^{i\alpha} \phi(x)$. It is analogous to the $U_{PQ}(1)$ 
symmetry of Peccei and Quinn. It is spontaneously broken by the vacuum 
expectation value $\langle \phi \rangle = v_a e^{i\alpha}$. The associated 
Nambu-Goldstone boson is the axion. The last term in Eq.~(\ref{vi1}) 
represents the non-perturbative QCD effects that give the axion its mass 
$m_a$. We have $m_a=\sqrt{\eta}$ for $v_a\gg m_a$.

When $\eta=0$, the model has global string solutions. A straight global
string along the $\hat{z}$-axis is the static configuration:
\begin{equation}
\phi(\vec{x})=v_a f(\rho) e^{i\theta} 
\label{2.2}
\end{equation}
where $(z,\rho,\theta)$ are cylindrical coordinates and $f(\rho)$ is a 
function which goes from zero at $\rho=0$ to one at $\rho=\infty$ over a 
distance scale of order $\delta \equiv (\sqrt{\lambda} v_a)^{-1}$.  
$f(\rho)$ may be determined by solving the non-linear field equations 
derived from Eq.~(\ref{vi1}). The energy per unit length of the global 
string is
\begin{equation}
\mu \simeq  2\pi \int^L_\delta \rho d\rho \frac{1}{2} |\vec{\nabla}\phi|^2 = 
\pi v_a^2 \ln (\sqrt{\lambda}v_aL)
\label{2.3}
\end{equation}
where $L$ is a long-distance cutoff. Eq.~(2.3) neglects the energy per unit
length, of order $v_a^2$, associated with the core of the string.  For a 
network of strings with random directions, as would be present in the early 
universe, the infra-red cutoff $L$ is of order the distance between strings.

When $\eta\neq 0$, the model has domain walls. If $v_a\gg m_a$ we may 
set $\phi(x) = v_a e^{\frac{i}{v_a} a(x)}$ when restricting ourselves to 
low energy configurations. The corresponding effective Lagrangian is :
\begin{equation}
{\cal L}_a = \frac{1}{2}\partial_\mu a\partial^\mu a + m_a^2 v_a^2\left[\cos
\left(\frac{a}{v_a}\right)-1\right] .
\label{vi2}
\end{equation}
Its equation of motion has the well-known Sine-Gordon soliton solution :
\begin{equation}
\frac{a(y)}{v_a}= 2\pi n + 4 \tan^{-1}\exp(m_a y) 
\label{2.5}
\end{equation}
where $y$ is the coordinate perpendicular to the wall and $n$ is any integer.
Eq.~(\ref{2.5}) describes a domain wall which interpolates between 
neighboring vacua in the low energy effective theory (\ref{vi2}). In the 
original theory (\ref{vi1}), the domain wall interpolates between the unique 
vacuum back to that same vacuum going around the bottom of the Mexican hat
potential 
$V(\phi^\dagger\phi)=\frac{\lambda}{4}(\phi^\dagger\phi - v_a^2)^2$ once. 

The thickness of the wall is of order $m_a^{-1}$. The wall energy per unit 
surface is $\sigma= 8 m_a v_a^2$ in the toy model.  In reality the structure 
of axion domain walls is more complicated than in the toy model, mainly
because not only the axion field but also the neutral pion field is spatially
varying inside the wall \cite{Huang}. When this is taken into account, the 
(zero temperature) wall energy/surface is found to be: 
\begin{equation}
\sigma \simeq 4.2~f_\pi m_\pi f_a \simeq 9~m_a f_a^2\ 
\label{2.6}
\end{equation}
with $f_a \equiv v_a/N$.  For $N=1$, $f_a$ and $v_a$ are the same.  However, 
written in terms of $f_a$, Eq. (2.6) is valid for $N \neq 1$ also. 

When $m_a\neq 0$, each string becomes the boundary of a domain wall.
Indeed the presence of the string forces 
$\alpha(\vec{x})=\frac{1}{v_a} a(\vec{x})$ 
to vary by $2\pi$ when $\vec{x}$ goes around the string once and a domain 
wall is the minimum energy path that does this. Fig.~1 shows a cross-section 
of a wall bounded by a string.  Two different length scales are 
involved: the wall thickness $m_a^{-1}$ and the string core size 
$\delta = (\sqrt{\lambda} v_a)^{-1}$.  Walls bounded by string are of 
course topologically unstable.   

In the early universe, the axion is essentially massless from the PQ phase
transition at temperature of order $v_a$ to the QCD phase transition at
temperature of order $1$ GeV. During that epoch, axion strings are present as
topological defects. At first the strings are stuck in the plasma and are
stretched by the Hubble expansion.  However with time the plasma becomes 
more dilute and below temperature \cite{Harari}
\begin{equation}
T_* \sim 2~10^7 \mbox{GeV} \left({v_a \over 10^{12} \mbox{GeV}}\right)^2
\label{2.7}
\end{equation}
the strings move freely and decay efficiently into axions. Because this decay 
mechanism is very efficient, there is approximately one string per horizon 
from temperature $T_*$ to temperature $T_1\simeq 1$ GeV when the axion 
acquires mass. The decay of axion strings into axions  has been discussed 
extensively in the literature and is reviewed in subsection \ref{sd}.

At temperature $T_1$ each string becomes the boundary of a domain wall. 
Walls not bounded by string are also present. We want to obtain 
statistical properties of this network of walls and strings under the 
assumption that the axion field is randomly oriented from one horizon 
to the next just before the axion mass turns on at time $t_1$. In particular: 
What are the average densities of walls and of strings?  Are there walls of 
\underline{infinite} extent which are not cut up by any string?  What is 
the energy fraction in walls not cut up by string?

To address such questions, a cross-section of a finite but statistically 
significant volume of the universe near time $t_1$ was simulated in the 
following manner \cite{Paris}.  A 2-dim. hexagonal grid was constructed.  
Each hexagon represents a causally disconnected region.  The circle 
$\{\langle \phi\rangle = 
v_a e^{i\alpha}: - \pi<\alpha=\frac{a}{v_a} \leq + \pi\}$ 
is divided into three equal parts as shown in Fig.~2a. $\alpha=0$ is at 
the middle of part 1. $\alpha=\pi$ is at the boundary between parts 2 and 
3. Each hexagon is then randomly assigned a number 1, 2 or 3.  A hexagon 
to which \#1 is assigned is thought of as a horizon volume in which the 
average value of $\alpha$ is in part 1 of the circle, and so on. Thus 
there is a domain wall at each boundary between two hexagons one of which 
has been assigned \#2 and the other \#3.  Likewise there is an downward-going 
string at each vertex surrounded by hexagons to which \#1, \#2 and \#3 have 
been assigned in clockwise order, and a upward-going string at each vertex 
surrounded by hexagons to which \#1, \#2 and \#3 have been assigned in 
counter-clockwise order.  Fig.~2b shows some examples.  Fig.~3 shows a 
larger realization with the underlying grid removed. 

Fig.~3 shows that there are no infinite domain walls which are not cut up 
by any string.  The reason for this is simple.  An extended domain wall 
has some probability to be cut up by string in each successive horizon it 
traverses.  The probability that no string is encountered after traveling 
a distance $l$ along the wall decreases exponentially with $l$.  Fig.~3 also 
shows that finite domain walls which are not cut up by string, and which 
therefore close onto themselves like a sphere or a donut, are very rare.  
Indeed, the figure shows only one closed circle and a circle need not be a 
section of a sphere; it could be a section of a tube.  The average density 
of walls at time $t_1$ predicted by the random process described here is 
approximately 2/3 per horizon volume.  This is consistent with Fig.~3.

\section{The cold axion populations.}
\label{cap}

In this section, we discuss the various contributions to the cosmological 
energy density in the form of {\it cold} axions.  In addition to cold axions, 
there is a thermal axion population whose properties are derived by the usual 
application of statistical mechanics to the early, high temperature universe
\cite{KT}.  The contribution of thermal axions to the cosmological energy 
density is subdominant if $m_a \lesssim 10^{-2}$ eV.  We concern ourselves 
only with cold axions here.

We distinguish the following sources of cold axions:\\
1. vacuum realignment\\
\indent\hspace{.5cm}a. ``zero momentum'' mode\\
\indent\hspace{.5cm}b. higher momentum modes\\
2. axion string decay\\
3. axion domain wall decay\\
The basis for distinguishing the contribution from vacuum realignment 
from the other two is that it is present for any quasi-Nambu-Goldstone 
field regardless of the topological structures associated with that field.
Contributions 1a and lb are distinguished by whether the misalignment 
of the axion field from the CP conserving direction is in modes of
wavelength larger (1a) or shorter (1b) than the horizon size $t_1$ at
the moment the axion mass turns on.  Contribution 2 is from the 
decay of axion strings before the QCD phase transition.  Contribution 3 is 
from the decay of axion domain walls bounded by strings after the 
QCD phase transition.  In the past literature on the axion cosmological 
energy density, contributions 1a and 2 have been discussed in detail 
whereas contributions 1b and 3 have received much less attention.  This 
section presents a systematic analysis of all contributions.

\subsection{Vacuum realignment}
\label{ivm}

Let us track the axion field from the PQ phase transition, when $U_{PQ}(1)$ 
gets spontaneously broken and the axion appears as a Nambu-Goldstone boson, 
to the QCD phase transition when the axion acquires mass.  For clarity of 
presentation we consider in this subsection a large region which happens to 
be free of axion strings.  Although such a region is rare in practice it may 
exist in principle.  In it, the contribution to the cosmological axion energy 
density from string and domain wall decay vanishes but that from vacuum 
realignment does not.  In more typical regions where strings are present, 
the contributions from string and domain wall decay should simply be added 
to that from vacuum realignment.

In the radiation dominated era under consideration, the space-time metric is 
given by:
\begin{equation}
- ds^2 = - dt^2 + R^2 (t)~d\vec x\cdot d\vec x
\label{3.1}
\end{equation}
where $t$ is the age of the universe, $\vec x$ are co-moving coordinates
and $R(t) \sim \sqrt{t}$ is the scale factor.  The axion field $a(x)$
satisfies:
\begin{eqnarray}
\left( D_\mu D^\mu + m_a^2 (t)\right) a(x)
= \left(\partial_t^2 + 3 \frac{\dot R}{R} \partial_t - \frac{1}{R^2}
\nabla_x^2 + m_a^2(t)\right) a(x) = 0
\label{3.2}
\end{eqnarray}
where $m_a (t) = m_a (T(t))$ is the time-dependent axion mass.  We will see 
that the axion mass is unimportant as long as $m_a(t) \ll 1/t$, a condition 
which is satisfied throughout except at the very end when $t$ approaches 
$t_1$.  The solution of Eq.~(\ref{3.2}) is a linear superposition of 
eigenmodes with definite co-moving wavevector $\vec k$:
\begin{equation}
a (\vec x, t) = \int d^3 k~~a(\vec k, t)~e^{i\vec k\cdot\vec
x}
\label{3.3}
\end{equation}
where the $a(\vec k, t)$ satisfy:
\begin{equation}
\left( \partial_t^2 + {3\over 2t} \partial_t + {k^2\over R^2} + m_a^2
(t)\right) a(\vec k, t) = 0\ .
\label{3.4}
\end{equation}
Eqs.~(\ref{3.1}) and (\ref{3.3}) show that the wavelength $\lambda (t)
= {2\pi\over k} R(t)$ 
of each mode is stretched by the Hubble expansion.  There are
two qualitatively different regimes in the evolution of each mode
depending on whether its wavelength is outside $(\lambda (t) > t)$ or
inside $(\lambda (t) < t)$ the horizon.

For $\lambda (t) \gg t$, only the first two terms in Eq.~(\ref{3.4}) 
are important and the most general solution is:
\begin{equation}
a(\vec k, t) = a_0 (\vec k) + a_{-1/2} (\vec k) t^{-1/2}\ .
\label{3.5}
\end{equation}
Thus, for wavelengths larger than the horizon, each mode goes to
a constant; the axion field is so-called ``frozen by causality''.

For $\lambda (t) \ll t$, the first three terms in Eq.~(\ref{3.4}) are 
important.  Let $a(\vec k, t) = R^{-{3 \over 2}}(t) \psi(\vec k, t)$.
Then
\begin{equation}
\left( \partial_t^2 + \omega^2(t)\right) \psi(\vec k, t) = 0 \,
\end{equation}
where
\begin{equation}
\omega^2(t) = m_a^2(t) + {k^2\over R^2 (t)} + {3\over 16t^2} \simeq
{k^2\over R^2(t)}\ .
\end{equation}
Since $\dot\omega \ll \omega^2$, this regime is characterized by 
the adiabatic invariant $\psi_0^2(\vec k,t)\omega(t)$, where
$\psi_0(\vec k,t)$  is the oscillation amplitude of $\psi(\vec k,t)$.
Hence the most general solution is:
\begin{equation}
a (\vec k, t) = {A\over R(t)} \cos \left( \int^t dt^\prime \omega
(t^\prime) \right)~~~~~\ .
\label{3.6}
\end{equation}
The energy density and the number density behave respectively as 
$\rho_{a,\vec k} \sim {A^2 w^2\over R^2(t)} \sim{1\over R^4 (t)}$ 
and $n_{a,\vec k} \sim {1\over \omega} \rho_{a,\vec k}
\sim {1\over R^3 (t)},$ indicating that the number of axions in each 
mode is conserved.  This is as expected because the expansion of the 
universe is adiabatic for $\lambda (t) t \ll 1$.

Let us define ${dn_a\over dw} (\omega, t)$ to be the number density, in
physical and frequency space, of axions with wavelength 
$\lambda=\frac{2\pi}{\omega}$, for $\omega >t^{-1}$.
The axion number density in physical space is thus:
\begin{equation}
n_a (t) = \int_{t^{-1}} d\omega~{dn_a\over dw} (\omega, t)\ ,
\label{3.8}
\end{equation}
whereas the axion energy density is:
\begin{equation}
\rho_a (t) = \int_{t^{-1}} d\omega {dn_a\over d\omega} (\omega, t)
\omega\ .
\label{3.9}
\end{equation}
Under the Hubble expansion axion energies redshift according to 
$\omega^\prime = \omega {R\over R^\prime}$, and volume elements expand 
according to 
$\Delta V^\prime = \Delta V\left({R^\prime\over R}\right)^3$, whereas
the number of axions is conserved mode by mode. Hence:
\begin{equation}
{dn_a\over d\omega} (\omega, t) = \left({R^\prime\over R}\right)^2
{dn_a\over d\omega} (\omega {R\over R^\prime}, t^\prime)\ .
\label{3.10}
\end{equation}
Moreover, the size of ${dn_a\over d\omega}$ for $\omega \sim {1\over t}$
is determined in order of magnitude by the fact that the axion field 
typically varies by $f_a$ from one horizon to the next.  Thus:
\begin{equation}
\left.\omega {dn_a\over d\omega} (\omega, t) \Delta\omega \right|_{\omega
\sim\Delta\omega\sim \frac{1}{t}}
\sim {dn_a\over d\omega} 
\left({1\over t}, t\right) \left({1\over
t}\right)^2 \sim {1\over 2} (\vec\nabla a)^2 \sim {1\over 2} {f_a^2\over
t^2}\ .
\label{3.11}
\end{equation}
{}From Eqs.~(\ref{3.10}) and (\ref{3.11}), and $R \sim \sqrt{t}$, we have:
\begin{equation}
{dn_a\over d\omega} (\omega, t) \sim {f_a^2\over 2t^2\omega^2}\ .
\label{3.12}
\end{equation}
Eq.~(\ref{3.12}) holds until the moment the axion acquires mass during the QCD
phase transition.  The critical time is when $m_a(t)$ is of order
$t^{-1}$.  Let us define $t_1$:
\begin{equation}
m_a (t_1) t_1 = 1 \ .
\label{3.13}
\end{equation}
$m_a(T)$ was obtained \cite{vacmis} from a calculation of the effects of QCD
instantons at high temperature \cite{highT}:  
\begin{equation}
m_a(T) \simeq 4 \cdot 10^{-9} {\rm eV} \left(\frac{10^{12}\rm GeV}{f_a}\right)
\left(\frac{\rm GeV}{T}\right)^4
\end{equation}
when $T$ is near $1$ GeV. The relation between $T$ and $t$ follows as usual
from
\begin{equation}
H^2=\left(\frac{1}{2t}\right)^2=\frac{8\pi G}{3} \rho = \frac{8\pi G}{3}
\cdot \frac{\pi^2}{30} {\cal N} T^4
\end{equation}
where ${\cal N}$ is the effective number of thermal degrees of freedom.  
${\cal N}$ is changing near 1 GeV temperature from a value near 60, valid above 
the quark-hadron phase transition, to a value of order 30 below that transition.
Using ${\cal N} \simeq 60$, one has
\begin{equation}
m_a(t) \simeq 0.7~10^{20} \frac{1}{\rm sec}\left(\frac{t}{\rm sec}\right)^2 
\left(\frac{10^{12}\rm GeV}{f_a}\right)  \ ,
\label{3.16n}
\end{equation}
which implies:
\begin{equation}
t_1 \simeq 2\cdot 10^{-7} \mbox{sec} 
\left({f_a\over 10^{12} \mbox{GeV}}\right)^{1/3} \ .
\label{3.14}
\end{equation}
The corresponding temperature is:
\begin{equation}
T_1 \simeq 1 \mbox{GeV} 
\left({10^{12} \mbox{GeV}\over f_a}\right)^{1/6}.
\label{3.15}
\end{equation}
We will make the usual assumption \cite{vacmis} that the changes in the 
axion mass are adiabatic [i.e. ${1\over m_a(t)} {dm_a\over dt} \ll m_a(t)$]
after $t_1$ and that the axion to entropy ratio is constant from $t_1$ till 
the present.  Various ways in which this assumption may be violated are 
discussed in the papers of ref.~\cite{entrop}.  We also assume that the 
axions do not convert to some other light axion-like particles as discussed 
in ref.~\cite{Hill}.

The above discussion neglects the non-linear terms associated with the 
self-couplings of the axion.  In general, such non-linear terms may cause 
the higher momentum modes to become populated due to spinodal instabilities 
and parametric resonance.  However, in the case of the axion field, such 
effects are negligible \cite{vacmis,Singh}.

We are now ready to discuss the vacuum realignment contributions to the
cosmological axion energy density.

\subsubsection{Zero momentum mode}

This contribution is due to the fact that, at time $t_1$, the axion field 
averaged over distances less than $t_1$ has in each horizon volume a random 
value between $-\pi f_a$ and $+\pi f_a$, whereas the CP conserving, and
minimum energy density, value is $a=0$.  The average energy density in this 
``zero momentum mode" at time $t_1$ is of order:
\begin{equation}
\rho_a^{vac,0} (t_1) \sim {1\over 2} m_a^2 (t_1) f_a^2\ .
\label{3.16}
\end{equation}
The corresponding average axion number density is:
\begin{eqnarray}
n_a^{vac,0} (t_1) = {1\over m_a(t_1)} \rho_a^{vac,0}
(t_1) \sim {f_a^2\over 2 t_1}\ .
\label{3.17}
\end{eqnarray}
Since $m_a(t)$ is assumed to change adiabatically after $t_1$, the number
of axions is conserved after that time.  Hence:
\begin{equation}
n_a^{vac,0} (t) \sim {1\over 2} {f_a^2\over t_1} \left({R_1\over
R}\right)^3\ .
\label{3.18}
\end{equation}
The axions in this population are non-relativistic and contribute $m_a$
each to the energy.

\subsubsection{Higher momentum modes}

This contribution is due to the fact that the axion field has wiggles
about its average value inside each horizon volume.  The axion number
density and spectrum associated with these wiggles is given in 
Eq.~(\ref{3.12}).  Integrating over $\omega > t^{-1}$, we find:
\begin{equation}
n_a^{vac,1} (t) = n_a^{vac,1} (t_1) \left({R_1\over R}\right)^3 \sim
{1\over 2} {f_a^2\over t_1} \left({R_1\over R}\right)^3
\label{3.19}
\end{equation}
for the contribution from vacuum realignment associated with
higher momentum modes.  The bulk of these axions are non-relativistic
after time $t_1$ and hence each axion contributes $m_a$ to the energy.  
Note that $n_a^{vac,0} (t)$ and $n_a^{vac,1} (t)$ have the same order 
of magnitude.

\subsection{String decay}
\label{sd}

The contribution from axions which were produced in the decay of axion 
strings has been discussed extensively in the literature.  We describe 
it here for the purpose of completeness and clarity.  There is still 
disagreement about the size of this contribution.  We will go rapidly 
over those points which are not controversial, but indicate explicitly 
those which are.

Axion strings have energy per unit length:
\begin{equation}
\mu \simeq \int_{L^{-1}}^{v_a} dk {\pi v_a^2\over k} = 
\pi v_a^2 \ln (v_a L)
\label{3.20}
\end{equation}
where $L$ is an infra-red cutoff. This is the same formula as 
Eq.~(\ref{2.3}) but with the integral written in wavevector space. 
In the early universe, when the strings have random directions relative 
to one another, $L$ is of order the distance between neighboring strings.  
The strings move relativistically and decay efficiently into axions.  
As a result there is approximately one \underbar {long} string per 
horizon at all times after the temperature has dropped below $T_*$ but
before the QCD phase transition.  By definition, a long string traverses
the horizon.  By intersecting each other with reconnection, the long 
strings produce closed string loops of size smaller than the horizon 
which decay efficiently into axions.

One long axion string per horizon implies the energy density:
\begin{equation}
\rho_{str} (t) \sim {\pi v_a^2\over t^2} \ln (t v_a)\ .
\label{3.21}
\end{equation}
We are interested in the {\it number} density of axions produced in the 
decay of axion strings because, as we will soon see, these axions 
become non-relativistic soon after time $t_1$ and hence contribute each 
$m_a$ to the energy.  The equations governing the number density 
$n_a^{str}(t)$ of axions produced in the decay of strings are:
\begin{equation}
{d\rho_{str}\over dt} = -2 H\rho_{str} - {d\rho_{str\rightarrow
a}\over dt}
\label{3.22}
\end{equation}
\begin{equation}
{dn_a^{str}\over dt} = - 3H n_a^{str} + {1\over \omega (t)}
{d\rho_{str\rightarrow a}\over dt}
\label{3.23}
\end{equation}
where $\omega (t)$ is defined by:
\begin{equation}
{1\over \omega (t)} = {1\over {d\rho_{str\rightarrow a}\over dt}} \int
{d\omega\over \omega} {d\rho_{str\rightarrow a}\over dt~d\omega}\ .
\label{3.24}
\end{equation}
In Eqs.~(\ref{3.22}-\ref{3.24}), ${d\rho_{str\rightarrow a}\over dt} (t)$ 
is the rate at which energy density gets converted from strings into 
axions at time $t$, whereas ${d\rho_{str\rightarrow a}\over dt~d\omega}$ 
is the spectrum of axions thus emitted.  The term $-2H \rho_{str} = + H
\rho_{str} - 3 H\rho_{str}$ in Eq.~(\ref{3.22}) takes account of the fact
that the Hubble expansion both stretches $(+ H\rho_{str})$ and dilutes
$(-3 H\rho_{str})$ long strings.  To obtain $n_a^{str} (t)$ from 
Eqs.~(\ref{3.21}-\ref{3.22}), we must know what $\omega (t)$ characterizes 
the spectrum ${d^2\rho_{str\rightarrow a}\over dt~d\omega}$ of axions
radiated by cosmic axion strings at time $t$.

The axions that are radiated at time $t$ are emitted by strings which
are bent over a distance scale of order $t$ and are relaxing to
lower energy configurations.  The main source is closed loops 
of size $t$. Two views have been put forth concerning the motion and 
the radiation spectrum of such loops.  One view \cite{Davis,stringA}
is that such a loop oscillates many times, with period of order $t$, 
before it has released its excess energy and that the spectrum of 
radiated axions is concentrated near ${2\pi\over t}$.  Let us call 
this scenario $A$.  A second view \cite{Harari,Hagmann} is that the 
loop releases its excess energy very quickly and that the spectrum 
of radiated axions 
${d\rho_{str\rightarrow a}\over dt~d\omega} \sim {1\over \omega}$ 
with a high energy cutoff of order $v_a$ and a low energy cutoff of order 
${2\pi\over t}$.  Let us call this scenario $B$.  In scenario $A$ one 
has $\omega (t) \sim {2\pi\over t}$ and hence, for $t < t_1$:
\begin{equation}
n_a^{str,A} (t) \sim {v_a^2\over t} \ln (v_a t) \ ,
\label{3.25}
\end{equation}
whereas in $B$ one has $\omega (t) \sim {2\pi\over t} \ln (v_a t)$ and 
hence:
\begin{equation}
n_a^{str,B} (t) \sim {v_a^2\over t}\ .
\label{3.26}
\end{equation}
Comparing Eqs.~(\ref{3.12}), (\ref{3.25}) and (\ref{3.26}), one sees
that in $A$ the contribution from string decay to the cosmological axion 
energy density dominates by a factor $\ln (v_a t) \sim 100$ over the 
contribution from vacuum realignment, whereas in $B$ the two
contributions are of the same order of magnitude.

We have carried out computer simulations \cite{Hagmann} of the motion 
and decay of axion strings with the purpose of deciding between the two 
possibilities.  In scenario $A$, the wavevector spectrum ${dE\over dk}$ 
of the energy stored in the axion field changes from 
${{\rm constant}\over k}$, characteristic of the string 
[see Eq.~(\ref{3.20})], to a spectrum peaked near $k\sim {2\pi\over t}$ 
after the string loop has decayed into axions.  In scenario $B$, 
${dE\over dk}$ does not qualitatively change.  To distinguish between 
the two possibilities, we defined the quantity:
\begin{equation}
N_{ax} \equiv \int dk {dE\over dk} {1\over k}\ ,
\label{3.27}
\end{equation}
and computed it as a function of time while integrating 
the equations of motion which follow from Lagrangian (\ref{vi1})
with $\eta = 0$.  In scenario $A$, $N_{ax}$ \underbar {increases} as 
the spectrum softens, whereas in $B$, $N_{ax}$ remains approximately 
constant.  We found in our simulations that $N_{ax}$ \underbar 
{decreases} by about 20\% during the decay of closed string loops.  
In the simulations, $\ln (v_a L)$ is in the range 4 to 6.  If scenario 
$A$ were correct, $N_{ax}$ should have increased by the factor 
$\ln (v_aL) - 1 = 3$ to 5 (depending upon the loop size $L$) 
during the decay of loops.  So our simulations support scenario $B$.  
Note that $N_{ax}$ is the quantity that directly measures the string 
contribution to the axion cosmological energy density.  An important 
caveat is that the simulations are done for values of $\ln (v_a L)$ 
much smaller than those, of order 60, which characterize axion 
strings in the early universe. 

We will assume henceforth that axion production by cosmic string decay 
is consistent with possibility $B$.  Hence: 
\begin{equation}
n_a^{str} (t) \sim {v_a^2\over t_1} \left({R_1\over R}\right)^3\ .
\label{3.28}
\end{equation}
It is of the same order of magnitude as the contributions 
[Eqs.~(\ref{3.18}) and (\ref{3.19})] from vacuum realignment.

\subsection{Domain wall decay}
\label{dwd}

During the QCD phase transition, when the axion acquires mass, the axion 
strings become the boundaries of axion domain walls.  Fig.~3 shows 
a cross-section of a piece of the universe at that time.  In 3 dimensions,
the domain walls bounded by string may be like pancakes or they may 
be long surfaces with holes.  Which type of configuration dominates 
matters little to the arguments given below.  It does matter that 
the universe is relatively free of domain walls which are not bounded 
by strings, i.e. domain walls which close onto themselves like spheres, 
donuts and so on.  We concluded in section \ref{wbs} that such domain 
walls are in fact very rare.  

Near $1$ GeV temperature, the time-dependent axion mass is given 
approximately by Eq.~(\ref{3.16n}). The wall energy per unit surface is
\begin{equation}
\sigma(t) \simeq 8 f_a^2 m_a(t) .
\label{3.32n}
\end{equation}
Since the chiral symmetry breaking phase transition has presumably not 
taken place yet, we do not use the zero-temperature result Eq.~(\ref{2.6}).
As illustrated in Fig.~1, the string at the boundary of a wall is embedded 
into the wall. Hence its infra-red cutoff $L$, in the sense of 
Eqs.~(\ref{2.3}) and (\ref{3.20}), is of order $m_a^{-1}$ and its energy 
per unit length is therefore
\begin{equation}
\mu \simeq \pi f_a^2 \ln (f_a/m_a)\ .
\label{3.29}
\end{equation}
The surface energy $E_\sigma$ of a typical (size $\sim t_1$) piece 
of wall bounded by string is $\sigma(t) t_1^2$ whereas the energy in 
the boundary is $E_\mu \sim \mu t_1$.  
There is a critical time $t_2$ when the ratio
\begin{eqnarray}
\frac{E_\sigma(t)}{E_\mu}\sim \frac{8 m_a(t)t_1}{\pi \ln (f_a/m_a)}
\label{3.30}
\end{eqnarray}
is of order one.
Using Eqs.~(\ref{3.16n}) and (\ref{3.14}), we (crudely) estimate :
\begin{equation}
t_2 \simeq  10^{-6} \mbox{sec} 
\left({f_a\over 10^{12} \mbox{GeV}}\right)^{1/3}
\label{3.35n}
\end{equation}
\begin{equation}
T_2 \simeq 600 \mbox{MeV} 
\left({10^{12} \mbox{GeV}\over f_a}\right)^{1/6}.
\label{3.36n}
\end{equation}
After $t_2$ the dynamics of the walls bounded by string is dominated by
the energy in the walls whereas before $t_2$ it is dominated by the
energy in the strings. A string attached to a wall is pulled 
by the wall's tension. For a straight string and flat wall, the
acceleration is:
\begin{eqnarray}
a_s(t) = {\sigma(t)\over\mu}  \simeq \frac{8 m_a(t)}{\pi \ln ( f_a/m_a)}
\simeq \frac{m_a(t)}{23} \simeq \frac{1}{t_1} \frac{m_a(t)}{m_a(t_2)}.
\label{3.31}
\end{eqnarray}
Therefore, after $t_2$, each string typically accelerates to relativistic 
speeds, in the direction of the wall to which is it attached, in less than 
a Hubble time.  We argue below that the string will then unzip the wall, 
and that the energy stored in the wall gets released in the form of barely 
relativistic axions.  However, before doing so, let us describe the 
competing decay mechanisms which have been discussed.

One mechanism is emission of gravitational waves \cite{Vil}.  A piece of 
wall bounded by string of size $\ell$, assuming for the moment that it is
long lived and that its motion is not damped by friction against the 
surrounding plasma, oscillates with frequency $\omega \sim \ell^{-1}$ and 
hence emits gravitational waves. The power $P$ may be estimated using the 
quadrupole formula:
\begin{equation}
P\sim - {d(\sigma\ell^2)\over dt} \sim G(\sigma\ell^4)^2 \omega^6 
\sim G\sigma^2\ell^2\ .
\label{3.32}
\end{equation}
Indeed, emission of gravitational waves by walls bounded by string of size 
$t_1$ is dominated by the wall contribution, rather than the string 
contribution, after time $t_2$.  Eq.~(3.39) implies the lifetime:
\begin{equation}
\tau_{\rm grav} \sim (G\sigma)^{-1} \simeq 3.10^3 \mbox{sec} 
\left({m_a\over 10^{-5} \mbox{eV}}\right)\ ,
\label{3.33}
\end{equation}
independently of size.  Note that a piece of wall bounded by string 
may self-intersect with reconnection producing two walls bounded by string 
which may again intersect with reconnection, and so on.  This fragmentation 
does not affect the lifetime (\ref{3.33}) although it shifts the spectrum 
of gravitational waves to higher frequencies.

Another energy dissipation mechanism is drag caused by the reflection of 
particles in the primordial plasma. The reflection coefficients between 
various particles and axion domain walls are discussed in ref.~\cite{Huang}.  
It was found there that axions which are not highly relativistic have 
reflection coefficient of order one and that the dominant friction 
experienced by the oscillating walls is on the fluid of cold axions that 
were produced by vacuum realignment and string decay.  Consider a bent 
wall of curvature radius of order $t_1$.  The tension in the wall
provides a driving force per unit surface of order $\sigma /t_1$ whereas 
reflection of cold axions produces a damping force per unit surface of order 
$2m_a \beta^2 n_a $ where $\beta$ is the velocity of the wall.  Adding the 
contributions from vacuum realignment [Eqs.~(\ref{3.17}) and (\ref{3.19})] 
and from string
decay [Eq.~(\ref{3.28})] we have 
$n_a (t) \sim {2f_a^2\over t_1} \left({t_1\over t}\right)^{3/2}$.  
This neglects the contribution $n_a^{dw} (t)$ of axions 
produced in the decay of the walls themselves but we will see below 
that $n_a^{dw} (t)$ is of the same order of magnitude as $n_a^{vac} 
(t)$ and $n_a^{str} (t)$.  Comparing the driving force with the damping 
force, one concludes that $\beta$ is of order one immediately after 
$t_1$ and hence that the rate of energy dissipation by friction is 
$P \sim - {d(\sigma\ell^2)\over dt} \sim m_a n_a (t) \ell^2$ which
implies:
\begin{equation}
- {d\ln\ell\over dt} \sim {1\over t_1} \left( {t_1\over t}\right)^{3/2}\ .
\label{3.34}
\end{equation}
Thus energy dissipation by friction is important just after the domain 
walls appear but this mechanism soon turns off as the universe becomes 
dilute.  

If emission of gravitational radiation and friction on the cold axion fluid 
were the only important dissipation mechanisms, the conclusion would 
be that a typical wall, whether or not bounded by string, loses a large 
fraction of its energy by friction immediately after $t_1$ and then lives 
till a time of order $\tau_{grav}$ when it decays into gravitons.

However we find it far more likely that the walls bounded by string decay 
into axions, long before $\tau_{grav}$.  If emission of gravitational waves 
were the dominant dissipation mechanism, a piece of wall bounded by string 
of size $\ell$, assuming for the moment that it does not reduce its size by 
self-intersections with reconnection, would oscillate on the order of 
$\frac{\tau_{\rm grav}}{\ell} \sim 10^{10}\left(\frac{t_1}{\ell}\right)
\left(\frac{m_a}{10^{-5}{\rm eV}}\right)^{4/3}$ times before decaying away.
This seems an implausibly large number considering that nothing forbids 
the wall from decaying into axions instead.  It is especially implausible 
when one considers that at each oscillation a wall bounded by string may 
self-intersect with reconnection producing two smaller walls bounded by 
string.   If the two pieces produced by an intersection with reconnection 
are typically of roughly equal size it takes only $\log_2(m_at_1)\sim 10$ 
iterations of this process for the size of the pieces to go down from 
$t_1$ to the thickness $m_a^{-1}$ of the walls.  Pieces of size $m_a^{-1}$ 
surely decay into axions because their dynamics is dominated by the energy 
of the strings and strings are known to decay into axions efficiently.  Even 
if the two pieces produced by a self-intersection with reconnection are 
typically of very different size, it takes at most $t_1m_a$ iterations for 
the pieces to go down in size to $m_a^{-1}$.  Hence for emission of 
gravitational waves to be the dominant decay mechanism, the probability $p$ 
of self-intersection with reconnection per oscillation can not be larger than 
$\frac{t_1^2 m_a}{\tau_{\rm grav}}\simeq 10^{-7}\left(\frac{10^{-5}{\rm
eV}}{m_a}\right)^{2/3}$. $p$ may reasonably be thought to be of order one.
It seems implausible that $p$ could be as small as $10^{-7}$.

In section \ref{sos} we describe our computer simulations of the motion and
decay of walls bounded by string.  In the simulations the walls decay
immediately, i.e. in a time of order their size divided by the speed of light. 
The average energy of the radiated axions is 
$\langle \omega_a \rangle\sim 7 m_a$ in the simulations.  However, 
as was already mentioned in the Introduction, the simulations are done for
$\ln(v_a/m_a)\simeq\ln(100)\simeq 4.6$, whereas in axion models of interest 
$\ln (v_a/m_a)\simeq 60$.  One reason this may affect the results is 
the following.  When massless the axion field is equivalent, by a duality 
transformation, to an anti-symmetric two index gauge field $A_{\mu\nu}$ 
\cite{Ramond}.  The axion string is a source for this field.  If the string 
accelerates with acceleration $a_s$ it emits radiation of frequency of order 
$a_s$.  Since $a_s\simeq \frac{m_a}{23}$ in axion models of interest 
[see Eq.~(\ref{3.31})], the emission of axion radiation is inhibited by the 
fact that $a_s$ is considerably less than $m_a$, the lowest possible 
frequency for axion radiation. In the simulations $a_s \simeq 0.6~m_a$.

At any rate, for the reasons given earlier, it is most plausible that the 
walls bounded by string decay into barely relativistic axions long before 
$\tau_{\rm grav}$.  Let $t_3$ be the time when the decay effectively takes 
place and let $\gamma \equiv \frac{\langle \omega_a \rangle}{m_a(t_3)}$ be
the average Lorentz $\gamma$ factor then of the axions produced. In 
section II, we estimated the density of walls at time $t_1$ to be of 
order $0.7$ per horizon volume.  Hence we estimate that between $t_1$ 
and $t_3$ the average energy density in walls is
\begin{equation}
\rho_{\rm d.w.}(t)\sim (0.7)(9)m_a(t)\frac{f_a^2}{t_1}\left(
\frac{R_1}{R}\right)^3 \, .
\label{3.41n}
\end{equation}
We used Eq.~(\ref{2.6}) and assumed that the energy in walls simply scales as
$m_a(t)$. After time $t_3$, the number density of axions produced in the decay
of walls bounded by strings is of order
\begin{equation}
n^{\rm d.w.}_a(t)\sim \frac{\rho_{\rm d.w.}(t_3)}{\langle \omega_a 
\rangle} \left(\frac{R_3}{R}\right)^3\sim \frac{6}{\gamma}
\frac{f_a^2}{t_1}\left(\frac{R_1}{R}\right)^3 \, .
\label{3.42n}
\end{equation}
Note that the dependence on $t_3$ drops out of our estimate of
$n^{\rm d.w.}_a$.  If we use the value $\gamma=7$ observed in our computer
simulations, we have
\begin{equation}
n^{\rm d.w.}_a(t)\sim\frac{f_a^2}{t_1}\left(\frac{R_1}{R}\right)^3 \, .
\label{3.43n}
\end{equation}
In that case the contributions from vacuum realignment 
[Eq.~(\ref{3.18}) and Eq.~(\ref{3.19})] and domain wall decay are of the 
same order of magnitude.  However, we will find in section IV that 
$\gamma$ rises approximately linearly with ln$(\sqrt{\lambda} v_a/m_a)$ 
over the range of ln$(\sqrt{\lambda} v_a/m_a)$ investigated in our numerical 
simulations.  If this behaviour is extrapolated all the way to 
ln$(\sqrt{\lambda} v_a/m_a) \simeq 60$, which is the value in axion models
of interest, then $\gamma \simeq 60$.  In that case the contribution 
from wall decay is subdominant relative to that from vacuum realignment. 

\subsection{The cold axion cosmological energy density}
\label{cacedr}

Adding the three contributions, and assuming scenario B for the string 
contribution, we estimate the present cosmological energy density in 
cold axions to be:
\begin{equation}
\rho_a (t_0) \sim 3 {f_a^2\over t_1} \left({R_1\over R_0}\right)^3 m_a\ .
\label{3.38}
\end{equation}
Following \cite{vacmis}, we may determine the ratio of scale factors 
$R_1/R_0$ by assuming the conservation of entropy from time $t_1$ till 
the present.  The number of effective thermal degrees of freedom at time 
$t_1$ is approximately ${\cal N}_1 \simeq 61.75$.  We do not include axions 
in this number because they are decoupled by then.  Let $t_4$ be a time 
(say $T_4 = 4$ MeV) after the pions and muons annihilated but before neutrinos 
decoupled.  The number of effective thermal degrees of freedom at time $t_4$ 
is ${\cal N}_4 = 10.75$ with electrons, photons and three species of neutrinos 
contributing.  Conservation of entropy implies
${\cal N}_1 T_1^3 R_1^3 = {\cal N}_4 T_4^3 R_4^3$.  The neutrinos decouple 
before $e^+e^-$ annihilation.  Therefore, as is well known, 
the present temperature $T_{\gamma 0} = 2.735~^0$K of the 
cosmic microwave background is related to $T_4$ by:  ${11\over 2} T_4^3 R_4^3 
= 2~T_{0\gamma}^3~R_0^3$.  Putting everything together we have
\begin{equation}
\left({R_1\over R_0}\right)^3 \simeq 0.063 \left({T_{\gamma, 0}\over T_1}
\right)^3\ .
\label{3.39}
\end{equation}
Combining Eqs.~(\ref{3.14}), (\ref{3.38}) and (\ref{3.39}),
\begin{equation}
\rho_a (t_0) \sim 10^{-29} {\mbox{gr}\over \mbox{cm}^3} 
\left({f_a\over 10^{12} \mbox{GeV}}\right)^{7/6} .
\label{3.40}
\end{equation}
Dividing by the critical density $\rho_c = {3H_0^2\over 8\pi G}$, we find :
\begin{equation}
\Omega_a \sim \left({f_a\over 10^{12} \mbox{GeV}}\right)^{7/6} 
\left({0.7\over h}\right)^2
\label{3.41}
\end{equation}
where $h$ is defined as usual by $H_0 = h~100 $km/s$\cdot$Mpc.

\subsection{Pop.~I and Pop.~II axions}
On the basis of the discussion in section IIIA-D we distinguish two kinds
of cold axions:\\

I) axions which were produced by vacuum realignment or string decay and
which were not hit by moving domain walls. They have typical momentum $\langle
p_I(t_1)\rangle \sim \frac{1}{t_1}$ at time $t_1$ because they are associated
with axion field configurations which are inhomogeneous on the horizon scale
at that time. Their velocity dispersion is of order:
\begin{equation}
\beta_I(t)\sim \frac{1}{m_at_1}\left(\frac{R_1}{R}\right)\simeq 3\cdot 
10^{-17}\left(\frac{10^{-5}{\rm eV}}{m_a}\right)^{5/6}\frac{R_0}{R} \ .
\label{3.48}
\end{equation}
Let us call these axions population I.\\

II) axions which were produced in the decay of domain walls. They have typical
momentum $\langle p_{II}(t_3)\rangle \sim \gamma m_a(t_3)$ at time $t_3$ when 
the walls effectively decay. Their velocity dispersion is of order:
\begin{equation}
\beta_{II}(t)\sim \gamma \frac{m_a(t_3)}{m_a}\frac{R_3}{R}\simeq 10^{-13}
\left(\ \gamma \frac{m_a(t_3)}{m_a}\frac{R_3}{R_1}\right) 
\left(\frac{10^{-5}{\rm eV}}{m_a}\right)^{1/6} \frac{R_0}{R} \, .
\label{3.49}
\end{equation}
The factor $q \equiv \gamma \frac{m_a(t_3)}{m_a}\frac{R_3}{R_1}$ parameterizes 
our ignorance of the wall decay process.  We expect $q$ to be of order one 
but with very large uncertainties.  Fortunately there is a lower bound on $q$ 
which follows from the fact that the time $t_3$ when the walls effectively 
decay must be after $t_2$ when the energy density in walls starts to exceed 
the energy density in strings. Using Eq.~(\ref{3.35n}), we find:
\begin{equation}
q=\frac{\gamma m_a(t_3)}{m_a}\frac{R_3}{R_1} > \frac{\gamma
m_a(t_2)}{m_a}\frac{R_2}{R_1} \simeq \frac{\gamma}{130}\left(
\frac{10^{-5}{\rm eV}}{m_a}\right)^{2/3} .
\label{5.6n}
\end{equation}  
Since our computer simulations suggest $\gamma$ is in the range 7 to 60, 
the axions of the second population (pop. II) have much larger velocity 
dispersion than the pop. I axions.\\ 

Note that there are axions which were produced by vacuum realignment or 
string decay but were hit by relativistically moving walls at some 
time between $t_1$ and $t_3$.  These axions are relativistic just after 
getting hit and therefore are part of pop.~II rather than pop.~I.

The fact that there are two populations of cold axions, with widely differing 
velocity dispersion, has interesting implications for the formation and 
evolution of axion miniclusters.  This is discussed in section \ref{am}.

Here we speculate that the primordial velocity dispersion of pop.~II axions 
may some day be measured in a cavity-type axion dark matter
detector \cite{experim}.  If a signal is found in such a detector, the energy 
spectrum of dark matter axions will be measured with great resolution. It has 
been pointed out that there are peaks in the spectrum \cite{peaks} because 
late infall produces distinct flows, each with a characteristic local velocity 
vector. These peaks are broadened by the primordial velocity dispersion, given 
in Eqs.~(\ref{3.48}) and (\ref{3.49}) for pop.~I and pop.~II axions 
respectively.  These equations give the velocity dispersion in intergalactic 
space.  When the axions fall onto the galaxy their density increases by a 
factor of order $10^3$ and hence, by Liouville's theorem, their velocity 
dispersion increases by a factor of order 10.  [Note that this increase is not 
isotropic in velocity space.  Typically the velocity dispersion is reduced in 
the direction longitudinal to the flow in the rest frame of the galaxy whereas 
it is increased in the two transverse directions.]  The energy dispersion 
measured on Earth is 
$\Delta E = m_a\beta \Delta \beta$ where $\beta \simeq 10^{-3}$ is the flow 
velocity in the rest frame of the Earth.  Hence we find 
$\Delta E_I\sim 3\cdot 10^{-19}\left(\frac{10^{-5}{\rm eV}}{m_a}\right)^{5/6}
m_a$ for pop.~I axions and 
$\Delta E_{II}\sim 10^{-15}~q~
\left(\frac{10^{-5}{\rm eV}}{m_a}\right)^{1/6}m_a$ for pop.~II.  The minimum 
time required to measure $\Delta E$ is $(\Delta E)^{-1}$.  This assumes ideal 
measurements and also that all sources of jitter in the signal not due to 
primordial velocity dispersion can be understood.  There is little hope of 
measuring the primordial velocity dispersion of pop.~I axions since
$(\Delta E_I)^{-1}\sim 10~{\rm years} \left(\frac{10^{-5}{\rm eV}}{m_a}
\right)^{1/6}$.  However 
$(\Delta E_{II})^{-1} \sim ~$day $q^{-1} \left(\frac{10^{-5}{\rm eV}}{m_a}
\right)^{5/6}$, and hence it is conceivable that the primordial velocity 
dispersion of pop.~II axions will be measured.  

\section{Computer Simulations}
\label{sos}

We have carried out an extensive program of 2D numerical simulations of
domain walls bounded by strings. The Lagrangian (2.1) in finite difference
form is
\begin{eqnarray}
L&=&\sum_{\vec{n} }
\left\{ \frac{1}{2} \left[ \left(\dot{\phi}_1(\vec{n},t)\right)^2 +
\left(\dot{\phi}_2(\vec{n},t)\right)^2 \right] - \sum_{j=1}^2\frac{1}{2}
\left[ \left(\phi_1(\vec{n}+\hat{j},t)-\phi_1(\vec{n},t)\right)^2
\right.\right.
\nonumber\\
&&\left.+ \left(\phi_2(\vec{n}+\hat{j},t)-\phi_2(\vec{n},t)\right)^2 \right]
-\frac{1}{4} \lambda \left[\left(\phi_1(\vec{n},t)\right)^2
+\left(\phi_2(\vec{n},t)\right)^2 -1\right]^2 \nonumber\\ &&\left.
\hspace{-5mm}
\phantom{\frac{1}{2}}
+ \eta \left(\phi_1(\vec{n},t)-1\right)
\right\}
\end{eqnarray}
where $\vec{n}$ labels the sites.  In these units, $v_a = 1$, the wall 
thickness is $1/m_a=1/\sqrt{\eta}$ and the core size is 
$\delta=1/\sqrt{\lambda}$.  The lattice constant is the unit of length.  
In the continuum limit, the dynamics depends upon a single critical 
parameter, namely $m_a \delta = m_a/\sqrt{\lambda}$.

Large two-dimensional grids $(\sim 4000\times 4000)$ were initialized with
a straight or curved domain wall, at rest or with angular momentum.  The
initial domain wall was obtained by overrelaxation, starting with the 
Sine-Gordon ansatz $\phi_1+i\phi_2={\rm exp}(4i\,{\rm tan^{-1}exp}(m_a y))$
inside a strip of length the distance $D$ between the string and anti-string, 
and the true vacuum $(\phi_1=1,\,\phi_2=0)$ outside. The string and 
anti-string cores were approximated by 
$\phi_1+i\phi_2=-{\rm tanh}(.583\,r/\delta)\,{\rm exp}(\mp i\theta )$, 
where $r$ and $\theta$ are polar coordinates about the core center, and 
held fixed during relaxation.  Stable domain walls were obtained for 
$1/(m_a \delta)\gtrsim 3$.  A first-order in time and second-order in 
space algorithm was used for the dynamical evolution, with time step
$dt=0.2$. The boundary conditions were periodic throughout and the total
energy was conserved to better than 1\%.  If the angular momentum was
nonzero, the initial $\dot{\phi}$ was obtained as the difference
of two relaxed wall configurations a small time step apart, divided by
that time step.  

The evolution of the domain wall was studied for various values
of $\sqrt{\lambda}/m_a$, the initial wall length $D$ and the initial 
velocity $v$ of the strings in the direction transverse to the wall, the 
string and anti-string going in opposite directions.  The strings  
attached to the wall are rapidly accelerated by the wall tension, the 
potential and gradient energies of the wall being converted into kinetic 
energy of the strings.  Fig.~4 shows the longitudinal velocity of the core as 
a function of time for the case $m_a^{-1} = 400, \sqrt{\lambda}/m_a = 10$
and $v=0$.   An important feature is the Lorentz contraction of the core. 
For reduced core sizes $\delta/\gamma_s \lesssim  5$, where $\gamma_s$ is 
the Lorentz factor associated with the speed of the string core, there is 
`scraping' of the core on the lattice accompanied by emission of spurious 
high frequency radiation. This artificial friction eventually balances the 
wall tension and leads to a terminal velocity. In our simulations we always 
ensured being in the continuum limit.

For a domain wall without rotation ($v=0$), the string cores meet head-on 
and go through each other.  Several oscillations of decreasing magnitude 
generally occur before annihilation.  For $\gamma_s \simeq 1$ the string 
and anti-string coalesce and annihilate one another.  For 
$\gamma_s \gtrsim 2$, the strings go through each other and regenerate 
a new wall of reduced length.  The relative oscillation amplitude 
decreases with decreasing collision velocity.  Fig.~5 shows the core
position as a function of time for the case $m_a^{-1} = 1000, 
\sqrt{\lambda}/m_a = 14.3, D = 2896$.  Fig.~6a shows the gradient,
kinetic, string potential, wall potential and total energies as
a function of time.  By `string (wall) potential' energy, we mean 
the energy associated with the third (fourth) term in Eq.~(4.1).  The 
total energy is accurately conserved in spite of the violent goings-on 
when the string and anti-string meet.

We also investigated the more generic case of a domain wall with angular 
momentum.  The strings are similarly accelerated by the wall but string 
and anti-string cores rotate around each other.  Fig. 6b shows the
gradient, kinetic, string potential, wall potential and total energies as 
a function of time for the case 
$m_a^{-1} = 500, \sqrt{\lambda}/m_a = 28.3, D = 2096$, and $v = 0.25$.  
The field is displayed in Fig.~7 at various time steps for the case 
$m_a^{-1} = 100, \lambda = 0.01, D = 524 $ and $v = 0.6$.  No 
oscillation is observed.  All energy is converted into axion radiation
during a single collapse.  Oscillations occur only when the angular
momentum is too small for the string and anti-string cores to miss each 
other.

We performed spectrum analysis of the energy stored in the $\phi$ field
as a function of time using standard Fourier techniques.  The 
two-dimensional Fourier transform is defined by
\begin{equation}
\tilde{f}(\vec{p})=\frac{1}{\sqrt{L_xL_y}}\sum_{\vec{n}} 
\exp\left[2i\pi\left(\frac{p_xn_x}{L_x}+\frac{p_yn_y}{L_y}\right)\right]
f(\vec{n})
\end{equation}
for $p_j=1,\cdots,L_j$ where $j=x,y$.  The dispersion law is:
\begin{equation}
\omega_p=\sqrt{2 \left(2- \cos\frac{2\pi p_x}{L_x}- \cos\frac{2\pi
p_y}{L_y}\right)+m_a^2}.
\end{equation}
Fig.~8 shows the power spectrum $dE/d\log \omega$ of the $\phi$ field
at various times during the decay of a rotating domain wall for the case 
$m_a^{-1} = 500, \lambda = 0.0032, D = 2096$, and $v = 0.25$.  Initially,
the spectrum is dominated by small wavevectors, $k\sim m_a$.  Such a 
spectrum is characteristic of the domain wall.  As the wall accelerates the 
string, the spectrum hardens until it becomes roughly $1/k$ with a long
wavelength cutoff of order the wall thickness $1/m_a$ and a short wavelength 
cutoff of order the reduced core size $\delta/\gamma_s$.  Such a spectrum is 
characteristic of the moving string.  In the case of the figure, the string 
and anti-string cores annihilate at about the time of the fourth frame 
(Fig. 8d). The spectrum remains approximately $1/k$ during the annihilation 
process and thereafter.

As discussed in section IIIC, $\gamma = \langle \omega_a \rangle/m_a$, the
average energy of the radiated axions in units of the axion mass, is the 
quantity which determines the wall decay contribution to the axion 
cosmological energy density [see Eq.~(3.43)].  Fig.~9 shows the time evolution 
of $\langle \omega \rangle/m_a$ for $m_a^{-1} = 500, v = 0.25, D = 2096$ and 
various values of $\lambda$.  By
definition,
\begin{equation}
\langle \omega \rangle = \sum_{\vec{p}} E_{\vec{p}}/
\sum_{\vec{p}} \frac{E_{\vec{p}}}{\omega_{\vec{p}}}
\end{equation}
where $E_{\vec{p}}$ is the gradient and kinetic energy stored in mode 
$\vec{p}$ of the field $\phi$.  After the domain wall has decayed into 
axions, $\langle \omega \rangle = \langle \omega_a \rangle$.  Fig.~10 
shows the time evolution of $\langle \omega \rangle/m_a$ for 
$m_a^{-1} = 500, \lambda = 0.0016, D = 2096$ and various values of $v$.  
For small $v$ and/or small $\lambda$, i.e when the angular momentum is 
low and the core size is big, the string cores meet head on
and the strings have one or more oscillations.  In that case, 
$\langle \omega_a \rangle/m_a \simeq 4$.  This is consistent with the 
value, $\langle \omega_a \rangle/m_a \simeq 3$, found by Nagasawa and 
Kawasaki \cite{Nagasawa}.  The low angular momentum regime, when the 
strings oscillate, is characterized by low energy of the radiated axions.  
We believe this regime to be less relevant to wall decay in the early 
universe because it seem unlikely that the angular momentum of a wall 
at the QCD epoch could be small enough for the strings to oscillate then, 
especially when one considers that the actual wall configurations are 
3 dimensional rather than 2D.  

In the more generic case when no oscillations occur, 
$\langle \omega_a \rangle/m_a \simeq 7$.  Moreover, we find that 
$\langle \omega_a \rangle/m_a$ depends upon the critical parameter 
$\sqrt{\lambda}/m_a$, increasing approximately as the logarithm of that 
quantity; see Fig.~9.  This is consistent with the time evolution of 
the energy spectrum, described in Fig.~8.  For a domain wall, 
$\langle \omega \rangle \sim m_a$ whereas for a moving string 
$\langle \omega \rangle \sim m_a \ln (\sqrt{\lambda} v_a \gamma_s/m_a)$.
Since we find the decay to proceed in two steps: 1) the wall energy is 
converted into string kinetic energy, and 2) the strings annihilate 
without qualitative change in the spectrum, the average energy of 
radiated axions is 
$\langle \omega_a \rangle \sim m_a \ln (\sqrt{\lambda} v_a/m_a)$.
Assuming this is a correct description of the decay process for 
$\sqrt{\lambda} v_a/m_a \sim 10^{26}$, then  
$\langle \omega_a \rangle/m_a \sim 60$ in axion models of interest.

Present technology is inadequate for 3D simulations with sufficient
resolution for the purposes we are interested in, such as obtaining 
reliable estimates of the factor $\gamma$.  However, lower resolution 
3D and 2D simulations of string-wall networks were carried out in 
ref. \cite{Spergel}.  They showed qualitatively similar behaviour in 
2 and 3 dimensions.

\section{Axion miniclusters}
\label{am}

We saw in section III that the axion fluid is inhomogeneous with 
${\delta\rho_a\over \rho_a} = {\cal O}(1)$ at the time of the QCD phase 
transition.  As will be discussed below, the streaming length of the 
pop.~I axions (those from vacuum realignment and string decay) is too short 
for all these inhomogeneities to get erased by free streaming before the 
time $t_{eq}$ of equality between matter and radiation.  $t_{eq}$ is  
when density perturbations start to grow by gravitational instability.  
At time $t_{eq}$, the ${\delta\rho_a\over \rho_a} = {\cal O}(1)$ 
inhomogeneities in the axion fluid promptly form gravitationally bound 
objects, called axion miniclusters \cite{Hogan,Kolb}.  Of course, axion 
miniclusters occur only if there is no inflation or if inflation occurs 
with reheat temperature above the phase transition where $U_{PQ}(1)$ is 
spontaneously broken.  

Axion miniclusters were discussed in the seminal papers of Hogan and Rees
\cite{Hogan}, and of Kolb and Tkachev \cite{Kolb}.  However, in these 
pioneering studies the role of domain walls, and hence of axions from 
domain wall decay, was neglected.  Also the estimates of the minicluster
mass have varied considerably from one another.  Hence we think it
appropriate to give our own analysis of this interesting phenomenon.

As described in section III, there are two populations of cold axions,
pop.~I and pop.~II, with velocity dispersions given by Eqs.(\ref{3.48}) and 
(\ref{3.49}) respectively.  Both populations are inhomogeneous with 
${\delta\rho_a\over \rho_a} = {\cal O}(1)$ at the time of the QCD phase 
transition.  The free streaming length from time $t_1$ to $t_{eq}$ is:
\begin{eqnarray}
\ell_f = R(t_{eq}) \int_{t_1}^{t_{eq}} dt {\beta (t)\over R
(t)}
\simeq \beta (t_1) \sqrt{t_1 t_{eq}} \ln \left( {t_{eq}\over
t_1}\right) \,.
\label{6.1}
\end{eqnarray}
For the sake of definiteness, we assume henceforth $\Omega_0 = 1$ and 
$H_0 = 60$ km/s.Mpc, in which case $T_{eq} = 2$ eV.  The free streaming 
length should be compared with the size 
\begin{equation}
\ell_{mc}\sim t_1 {R_{eq}\over R_1} \simeq \sqrt{t_1 t_{eq}} \simeq
0.7\cdot 10^{13}{\rm cm} \left(\frac{10^{-5}{\rm eV}}{m_a}\right)^{1/6}
\end{equation} 
of axion inhomogeneities at time $t_{eq}$.  Using Eq.(\ref{3.48}) we find 
for pop.~I:
\begin{equation}
\frac{\ell_{f,I}}{\ell_{mc}} \simeq {1 \over t_1 m_a} \ln \left( 
{t_{eq}\over t_1}\right) \simeq 1.6\cdot 10^{-2} \left(\frac{10^{-5}
{\rm eV}}{m_a}\right)^{2/3} .
\label{6.2}
\end{equation}
Hence, in the axion mass range of interest, pop.~I axions
do not homogenize.  At $t_{eq}$ many pop.~I axions condense 
into miniclusters since they have ${\delta\rho\over\rho} = {\cal O}(1)$ 
upon arrival. The typical size of axion miniclusters  is $\ell_{mc}$
and their typical mass is:
\begin{equation}
M_{mc} \sim \frac{1}{4}~\rho_a (t_{eq}) {4\pi\over 3} \left( {\ell_{mc}\over
2}\right)^3 \sim 0.7~\cdot 10^{-13} M_\odot \left(\frac{10^{-5}
{\rm eV}}{m_a}\right)^{5/3} .
\label{6.5}
\end{equation}
We assumed that half the cold axions are pop.~I and that half of those form
miniclusters, hence the factor $1/4$.  We used Eq.~(3.48) to estimate
$\rho_a (t_{eq})$.

Using Eq.~(\ref{3.49}), we find for pop.~II:
\begin{equation}
\frac{\ell_{f,II}}{\ell_{mc}} \sim q \ln \left(
{t_{eq}\over t_3}\right) \simeq 40\, q .
\label{5.5n}
\end{equation}
Using Eq.~(3.51) and the range $\gamma \sim $ 7 to 60, suggested by
our numerical simulations, we conclude that pop.~II axions do homogenize and
hence that the axion energy density has a smooth component at $t_{eq}$.

At time t, pop.~II axions get gravitationally bound to the nearest 
minicluster if it is less than distance $d \sim GM_{mc}/\beta_{II}(t)^2$ away.
Since most are at a distance $d(t)\sim \ell_{mc}\frac{T_{eq}}{T(t)}$, the bulk 
of pop.~II axions become gravitationally bound to miniclusters when the
temperature has dropped to:
\begin{equation}
T_{II}\sim 4 \cdot 10^{-3} \frac{T_{eq}}{q^2} ,
\end{equation}
assuming $\Omega_a\sim 1$. Thus we arrive at a qualitative picture of
miniclusters before galaxy formation as having an inner core of pop.~I axions
and a fluffy outer envelope of pop.~II axions. The inner core has size
$\ell_{mc}$ and density
\begin{equation}
\rho_{mc} \sim 10^{-18} {\rm gr\over \mbox{cm}^3}\ .
\label{6.7}
\end{equation}
If $T_{II}>T_0$ (i.e. $q\mbox{ \raisebox{-1mm}{$\stackrel{<}{\sim}$} }
6 $), the outer envelope has roughly the same mass as the core but size
\begin{equation}
\ell_{mc}'\sim~~ 2~q^2~10^{15} {\rm cm} 
\end{equation}
and density 
\begin{equation}
\rho'_{mc}\sim~~q^{-6}~~ 10^{-25} \frac{\rm gr}{cm^3} ~~~~.
\end{equation}
If $T_{II}<T_0$, then the envelope is correspondingly less massive and less 
dense.  When a minicluster falls onto a galaxy, tidal forces of the galaxy 
are apt to destroy it.  We find below that the outer envelopes almost surely 
get pulled off whereas the inner cores barely survive.

When a minicluster passes by an object of mass $M$ with
impact parameter $b$ and velocity $v$, the internal energy per unit mass
$\Delta E$ given to the minicluster by the tidal gravitational forces from
that object is of order \cite{Hogan}
\begin{equation}
\Delta E \sim {G^2 M^2\ell_{mc}^2\over b^4 \beta^2}
\label{6.9}
\end{equation}
whereas the binding energy per unit mass of the minicluster 
$E\sim G~\rho_{mc} \ell_{mc}^2$.  If the minicluster travels a length
$\ell = \beta t$ through a region where objects of mass $M$ have density
$n$, the relative increase in internal energy is:
\begin{equation}
{\Delta E\over E} \sim {G\rho_M^2 t^2\over \rho_{mc}}\ ,
\label{6.10}
\end{equation}
where $\rho_M = Mn$.  Eq.~(\ref{6.10}) follows from the fact that  
$\Delta E$ is dominated by the closest encounter and the latter has 
impact parameter $b_{min}$:  $\pi b_{min}^2 n\ell = 0(1)$.  Note that 
${\Delta E\over E}$ is independent of $M$.  The effect upon a 
minicluster of falling \underbar {once} with velocity 
$\beta \simeq 10^{-3} $ through 
the inner halo $(r < 10~$kpc$)$ of our galaxy where 
$\rho_M \sim 10^{-24}$gr$/\mbox{cm}^3$ is:
\begin{eqnarray}
{\Delta E\over E} \sim {\rho_M\over \rho_{mc}} &\sim
10^{-6}&~~~~~ \mbox{ for the inner core}\nonumber \\
&\sim 10~~q^{6}&~~~~ \mbox{ for the outer envelope}
\label{6.11}
\end{eqnarray}
Since we expect $q$ to be of order one, it seems very likely that
the envelope gets pulled off after one or more crossings of the inner
parts of the galaxy.  This result is reassuring for direct searches of 
dark matter axions on Earth since it implies that a smooth component of 
dark matter axions with density of order $\rho_{halo}$  permeates us
whether or not there is inflation after the Peccei-Quinn phase transition. 
If there were no pop.~II axions one might fear that nearly all axions are 
in axion miniclusters.

This result can also be obtained by considering the tidal radius 
of a mini-cluster at our location in the galaxy:
\begin{equation}
R_{tidal} \sim ( { M_{mc} \over 3 M_{gal} } )^{1/3} D \sim
8~\cdot 10^{13} {\rm cm} \left(\frac{10^{-5}{\rm eV}}{m_a}\right)^{5/9}
\end{equation}
where $D \simeq 8.5$ kpc is our distance from the galactic center and 
$M_{gal} \simeq 10^{12} M_\odot$ is the mass of the galaxy interior
to us.  $R_{tidal}$ is larger than the inner core size $\ell_{mc}$
but smaller than the outer envelope $\ell_{mc}'$.

A minicluster inner core which has spent most of its life in the central
part of our galaxy only barely survived since ${\Delta E\over
E} \sim 10^{-2}$ in that case.  If most minicluster inner cores
have survived more or less intact, the direct encounter of a minicluster 
with Earth would still be quite rare, happening only every $10^4$ years or so.  
The encounter would last for about 3 days during which the local axion
density would increase by a factor of order $10^6$.

\section{THE THIRD SOLUTION TO THE AXION DOMAIN WALL PROBLEM}
\label{adwp}

In this section we consider axion models in which the number $N$ of 
degenerate vacua at the bottom of the Mexican hat potential is larger 
than one.  $N$ is an integer given by
\begin{equation}
\mbox{Tr}(Q_{PQ}Q^a_cQ^b_c)=N\delta^{ab} .
\end{equation}
Here $Q_{PQ}$ is the Peccei-Quinn charge, $Q_c^a~(a=1...8)$ are the 
color charges and Tr represents a sum over all the left-handed Weyl 
spinors in the model.  It is easy to construct models for an arbitrary 
value of $N$.  

There are $N$ distinct degenerate equally spaced vacua at the bottom of 
the Mexican hat potential \cite{degen}.  The vacua are related by an 
exact spontaneously broken $Z(N)$ symmetry. A domain wall is a transition 
in space from any one vacuum to a distinct neighboring vacuum.  A string 
is the boundary of $N$ walls.  If $N>1$ and there is no inflation after 
the Peccei-Quinn phase transition, axion models suffer from a domain wall 
problem because the energy density in walls ends up exceeding the energy 
density in matter and radiation.  A domain wall dominated universe is 
very much unlike our own.  The walls are gravitationally repulsive
\cite{repuls}.  The universe is divided into rapidly expanding bubbles 
with walls at the boundaries and islands of matter in the middles.  
When averaged on scales larger than the bubble size, the universe 
expands as $R \sim t^2$ \cite{Zel}.  At least in the case of axion 
walls, the islands are far too small for one island to be identified 
with our visible universe.

The axion domain wall problem can be solved 1) by having inflation 
with reheating temperature less than the PQ phase transition temperature,
or 2) by having $N=1$ as discussed in the other sections of this paper,
or 3) by having a small explicit breaking of the $Z(N)$ symmetry 
\cite{degen}.  The third solution is discussed in this section.  The 
symmetry breaking must lift completely the degeneracy of the vacuum and 
be large enough that the unique true vacuum takes over before the walls 
dominate the energy density. On the other hand, it must be small enough 
that the PQ mechanism still works.  The implications of biased discrete 
symmetry breaking have been discussed in a general manner in 
ref. \cite{Gel}.
 
We may, for our purposes, take the effective action density for the 
axion field to be:
\begin{equation}
{\cal L}_a = {1\over 2} \partial_\mu a \partial^\mu a + 
\frac{m_a^2 v_a^2}{N^2} \left[\cos \left({Na\over v_a}\right) -1\right]\ .
\label{5.1}
\end{equation}
The axion field $a(x)$ is cyclic with period $2\pi v_a$.  Thus 
Eq.~(\ref{5.1}) implies $N$ degenerate vacua.  The axion decay constant 
is $f_a \equiv {v_a\over N}$.  The domain wall is a transition in space
from a vacuum, say $a = 0$, to a neighbor vacuum, say
$a = \frac{2\pi v_a}{N}$.  The energy per unit surface is $\sigma = 
8 m_a f_a^2$.  The actual axion domain wall is only crudely 
described by Eq.~(\ref{5.1}) because it involves not only a rotation of 
the axion field but also of the quark-antiquark condensates.  However 
Eq.~(6.2) is adequate for the order of magnitude estimates we are 
interested in here.

Between the PQ and QCD phase transitions, axion strings are present and 
evolve as in the $N=1$ case.  At time $t_1$ each string becomes the 
boundary of $N$ walls.  For a brief period the walls lose energy by 
friction against the primordial plasma but this dissipation process 
is soon turned off as the plasma becomes dilute.

Causality implies that there is at least one domain wall per horizon on
average since each horizon volume picks a vacuum independently from its 
neighbor.  If there is of order one wall per horizon, the energy density 
in walls $\rho \sim {\sigma t^2\over t^3} = {\sigma\over t}$.  In that 
case, the energy in walls per co-moving volume 
$(E\sim \rho R^3 \sim \sqrt{t})$ 
increases with time.  This suggests that there is not more than 
approximately one wall per horizon because energy in walls has low 
entropy and hence increases only reluctantly.

With one wall per horizon, the fraction of the energy density in walls is
\begin{equation}
\Omega_{d.w.} (t) \sim {32\pi Gt\sigma\over 3} \sim 2.10^{-9} \left({10^{-5}
\mbox{eV}\over m_a}\right)^{4/3} {t\over t_1}\ .
\label{5.2}
\end{equation}
If this fraction reaches one, the universe becomes domain wall dominated 
and disaster strikes as described above.  To get rid of the walls, we add 
to the model a tiny $Z(N)$ breaking term which lifts the vacuum degeneracy 
completely, e.g.:
\begin{equation}
\delta V = - \xi (\varphi e^{-i\delta} + h.c.)
\label {5.3}
\end{equation}
where $\varphi$ is the PQ field with the ``Mexican hat'' potential, at 
the bottom of which $\varphi (x) = v_a~e^{ia(x)/v_a}$.  On the RHS 
of Eq.(6.2) an extra term appears:
\begin{equation}
\delta {\cal L}_a = 2 v_a\xi~\cos \left({a\over v_a} - \delta\right)\ .
\label{5.4}
\end{equation}
Now the unique true vacuum is the one for which $\mid \delta 
- {a\over v_a}\mid$ is smallest.  Its energy density is lowered by an 
amount of order $\xi v_a$ relative to the other quasi-vacua.  As a 
result, the walls at the boundary of a region in the true vacuum are 
subjected to an outward pressure of order $\xi v_a$.  Since the walls 
are typically a distance $t$ apart, the volume energy $\xi~v_a~t^3$ 
associated with the lifting of the vacuum degeneracy grows more rapidly 
than the energy $\sigma t^2$ in the walls.  At a time 
$\tau \sim {\sigma\over \xi v_a}$, the pressure favoring the 
true vacuum starts to dominate the wall dynamics and the true vacuum 
takes over, i.e. the walls disappear.  We require this to happen before 
the walls dominate the energy density.  Using Eqs.~(\ref{3.14}) and 
(\ref{5.2}), we obtain:
\begin{equation}
\tau \sim {\sigma\over \xi v_a} \lesssim 10^2~\mbox{sec}~
\left({m_a\over 10^{-5} \mbox{eV}}\right)\ .
\label{5.5}
\end{equation}
On the other hand $\xi$ is bounded from above by the requirement 
that $\delta V$ does not upset the PQ mechanism.  $\delta V$ shifts the 
minimum of the effective potential for the axion field, inducing a  
$\bar\theta \sim {\xi\over m_a^2 f_a}$.  The requirement that 
$\bar\theta < 10^{-10}$ implies:
\begin{equation}
\tau \gtrsim \mbox{sec} \left({10^{-5} \mbox{eV}\over m_a}\right)\ .
\label{5.7}
\end{equation}
Eqs.~(\ref{5.5}) and (\ref{5.7}) indicate that there is very little room in 
parameter space for this third solution to the axion domain wall problem, at 
least if $m_a \sim 10^{-5}$ eV.  But the third solution is not completely 
ruled out.

Soon after time $\tau$, a cross-section of a small fraction of the universe 
looks like Fig.~3 except that every string -- anti-string pair is connected 
by $N$ walls instead of just one wall.  This likely changes the mechanism by 
which the walls bounded by strings dissipate their energy since each string 
is being pulled in different directions by the $N$ walls to which it is 
attached.  

The fraction $r$ of wall energy that goes into non-relativistic or barely 
relativistic axions is at any rate severely restricted since $\tau$ is much 
earlier than the time $t_{eq} \sim 3~\cdot 10^{11}$ sec of equality between 
matter and radiation and, using Eqs.~(\ref{5.2}) and (\ref{5.7}),
\begin{eqnarray}
\Omega_{d.w.} (\tau) &\sim& 10^{-2} \left({10^{-5} \mbox{eV}\over m_a}\right)
{\tau\over\rm sec}
\gtrsim 10^{-2} \left({10^{-5} \mbox{eV}\over m_a}\right)^2\ .
\label{5.8}
\end{eqnarray}
The fraction of the cosmological energy density in axions from domain wall 
decay
\begin{equation}
\Omega_a^{d.w.} (t) = r~\Omega_{d.w.} (\tau) \left({t\over \tau}\right)^{1/2}
\label{5.9}
\end{equation}
should be less than about 1/2 when $t = t_{eq}$.  This implies the bound
\begin{eqnarray}
r \lesssim 10^{-4} \left({m_a\over 10^{-5} \mbox{eV}}\right) 
\left({\mbox{sec}\over\tau}\right)^{1/2}
\lesssim ~10^{-4} \left({m_a\over 10^{-5} \mbox{eV}}\right)^{3/2},
\label{5.10}
\end{eqnarray}
where we used Eq.~(\ref{5.7}) to obtain the second inequality.

However emission of gravitational waves is likely to be very effective 
because $\tau_{g.w.}$, given by Eq.~(3.40), is not very much larger than 
$\tau$.  Thus decay of the walls into gravitational radiation is in this 
case both necessary and likely.  The third solution therefore predicts a 
peak in the present spectrum of gravitational waves at the frequency
\begin{equation}
\omega_0 \sim {1\over\tau} \left({R_\tau\over R_0}\right) \sim 2.10^{-10} H_z
\left({\mbox{sec}\over \tau}\right)^{1/2}\ .
\label{5.11}
\end{equation}
The energy density in these gravitational waves is of order
\begin{equation}
\rho_{g.w.} (t_0)\sim 10^{-2} \rho_\gamma (t_0) \left({10^{-5} \mbox{eV}\over
m_a}\right) {\tau\over \mbox{sec}}\ ,
\label{5.12}
\end{equation}
where $\rho_\gamma (t_0)$ is the energy density in the cosmic microwave 
background.

\section{Conclusions.}
\label{con}

We have discussed the fate of the walls bounded by strings which appear
during the QCD phase transition in axion models with $N=1$ assuming the 
axion field is not homogenized by inflation.  We have argued that the main 
decay mechanism of these objects is radiation of barely relativistic axions 
and that the decay takes place during or soon after the QCD phase transition.

We presented the results of our computer simulations of the motion and
decay of walls bounded by strings.  In the simulations, the walls decay 
immediately, i.e. in a time scale of order the light travel time.  The 
computer simulations also provide an estimate of the average energy of 
the axions emitted in the decay of the walls: 
$\langle \omega_a \rangle \simeq 7 m_a$ when 
$\sqrt{\lambda}v_a/m_a \simeq 20$.
 
Because of restrictions on the available lattice sizes, the simulations are 
for values of $v_a/m_a$ of order 100, whereas in axion models of interest 
$v_a/m_a$ is of order $10^{26}$.  To address this shortcoming, we have 
investigated the dependence of $\langle \omega_a \rangle /m_a$ upon 
$\sqrt{\lambda}v_a/m_a$ and found that it increases approximately as the 
logarithm of this quantity.  This is because the decay process occurs in 
two steps: 1) wall energy converts into moving string energy because the 
wall accelerates the string, and the energy spectrum hardens accordingly; 
2) the strings annihilate into axions without qualitative change in the 
energy spectrum.  If this behaviour persists all the way to 
$\sqrt{\lambda}v_a/m_a \sim 10^{26}$, then 
$\langle \omega_a \rangle /m_a \sim 60$ for axion models of interest.  

To parameterize our ignorance of the decay process we introduced two 
parameters: the time $t_3$ when the decay effectively takes place
and the average Lorentz factor $\gamma = {\langle \omega_a \rangle /m_a}$ 
of the radiated axions at time $t_3$.  We found that in first 
approximation the cosmological energy density of axions from wall 
decay does not depend upon $t_3$ whereas it is inversely proportional 
to $\gamma$.  For $\gamma \sim 7$ the contribution from wall decay is 
of the same order of magnitude as that from vacuum realignment.  All 
three contributions - vacuum realignment, string decay and wall decay - 
were discussed in one unified treatment.

We discussed the velocity dispersion of axions from wall decay 
(pop.~II) and found that it is much larger, by a factor $10^3$ or so, 
than that of axions from vacuum realignment and from string decay 
(pop.~I).  This has interesting consequences for the formation and 
evolution of axion miniclusters.  We showed that the QCD horizon 
scale density perturbations in pop.II axions get erased by free 
streaming before the time $t_{eq}$ of equality between radiation 
and matter whereas those in pop.I axions do not.  Pop.~II axions
form an unclustered component of the axion cosmological energy 
density which guarantees that the signal in direct searches for 
axion dark matter is on all the time.

We speculated that the velocity dispersion of axions from wall decay 
may be large enough to be measured in a cavity detector of dark 
matter axions.  This would provide a direct experimental handle on 
some of the issues raised by axion cosmology, e.g. the question 
whether inflation occurred below or above the PQ phase transition.      

Finally we discussed the cosmology of axion models with $N>1$ in 
which the axion domain wall problem is solved by postulating 
a small U$_{PQ}$(1) breaking interaction which slightly lowers 
one of the $N$ vacua with respect to the others.  There is very 
little room in parameter space for this to work but it is a 
logical possibility.  We find that in this case the walls must
decay into gravitational radiation of frequency of order 
$10^{-10}$Hz.

\acknowledgments
We thank E.W. Kolb and E.P.S. Shellard for useful comments.  PS was 
the beneficiary of a Fellowship from the J.S. Guggenheim Memorial 
Foundation.  This research was also supported in part by DOE grant 
DE-FG05-86ER40272 at the University of Florida and by DOE grant 
W-7405-ENG-048 at Lawrence Livermore National Laboratory.

%
\begin{figure}[b]
\begin{center}
\vspace{1cm}
\includegraphics[width=14cm]{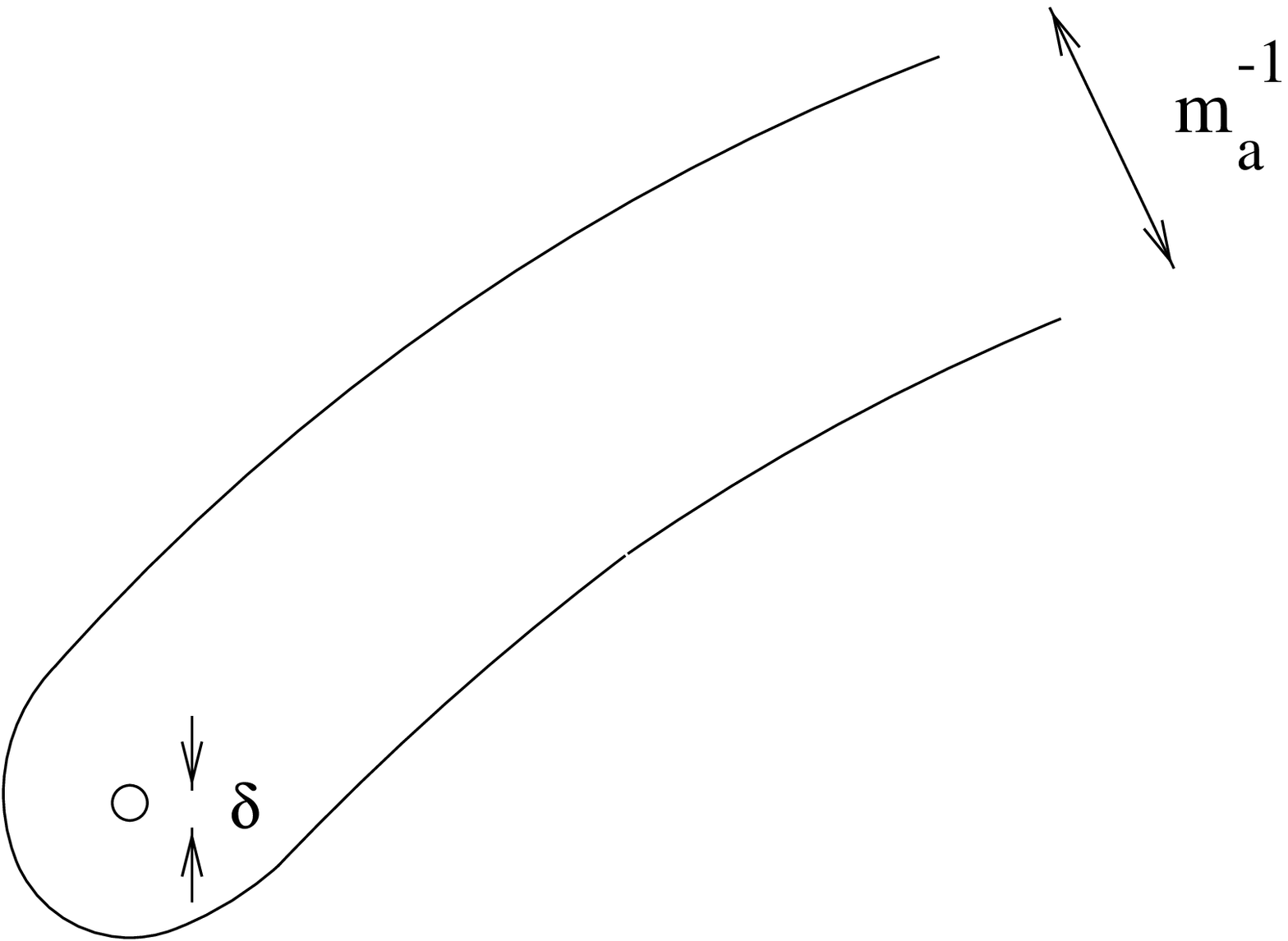}
\vspace{1cm}
\end{center}
\caption{Cross-section of a piece of wall bounded by a string.  The 
wall thickness is $m_a^{-1}$.  The string core size is
$\delta = 1/\sqrt{\lambda}v_a$.}
\end{figure}
\newpage
\begin{figure}
\begin{center}
\includegraphics[angle=-90,width=7cm]{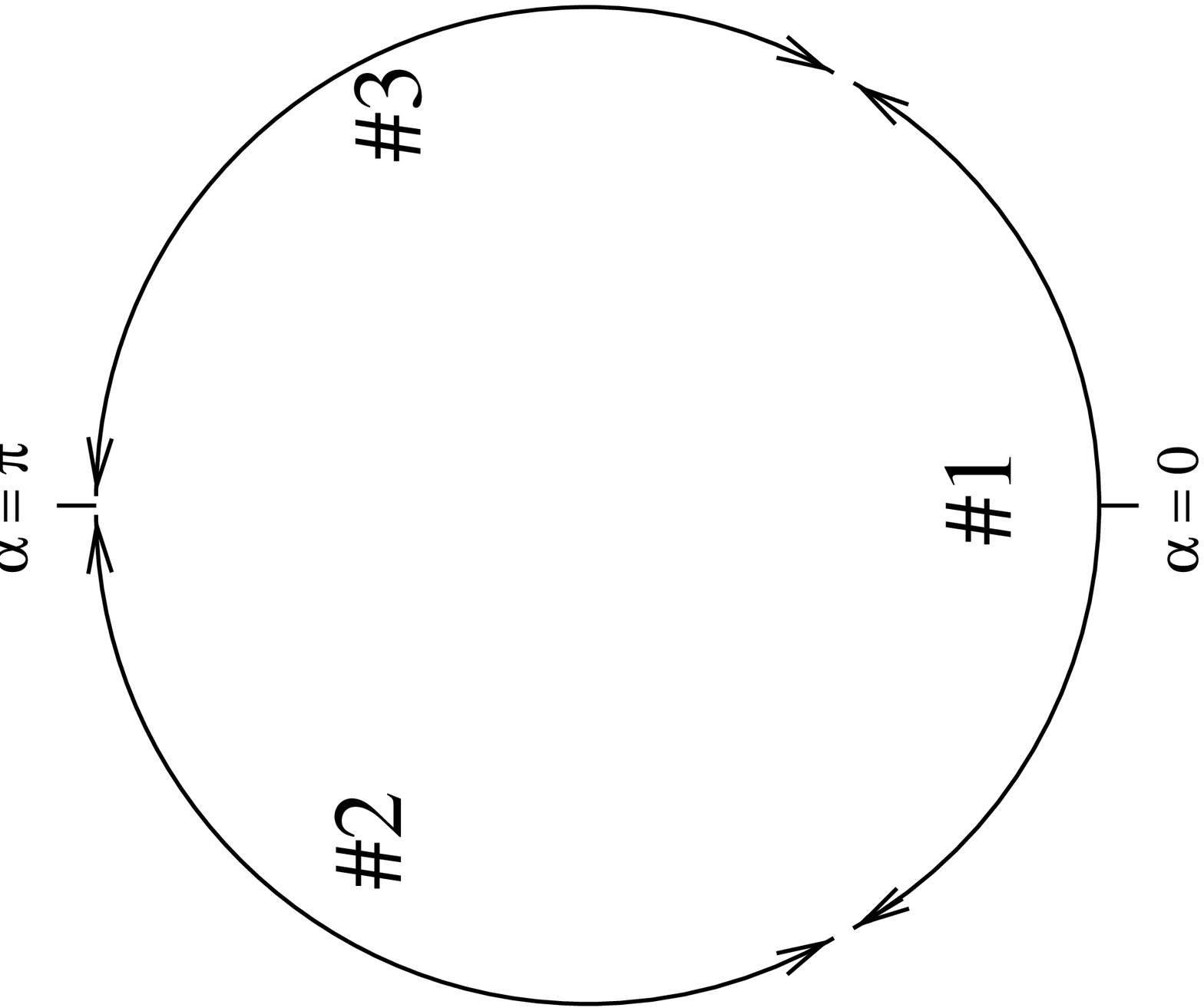}
\\[1mm]
{\bf (a)}
\end{center}
\begin{center}
\includegraphics[angle=-90,width=12cm]{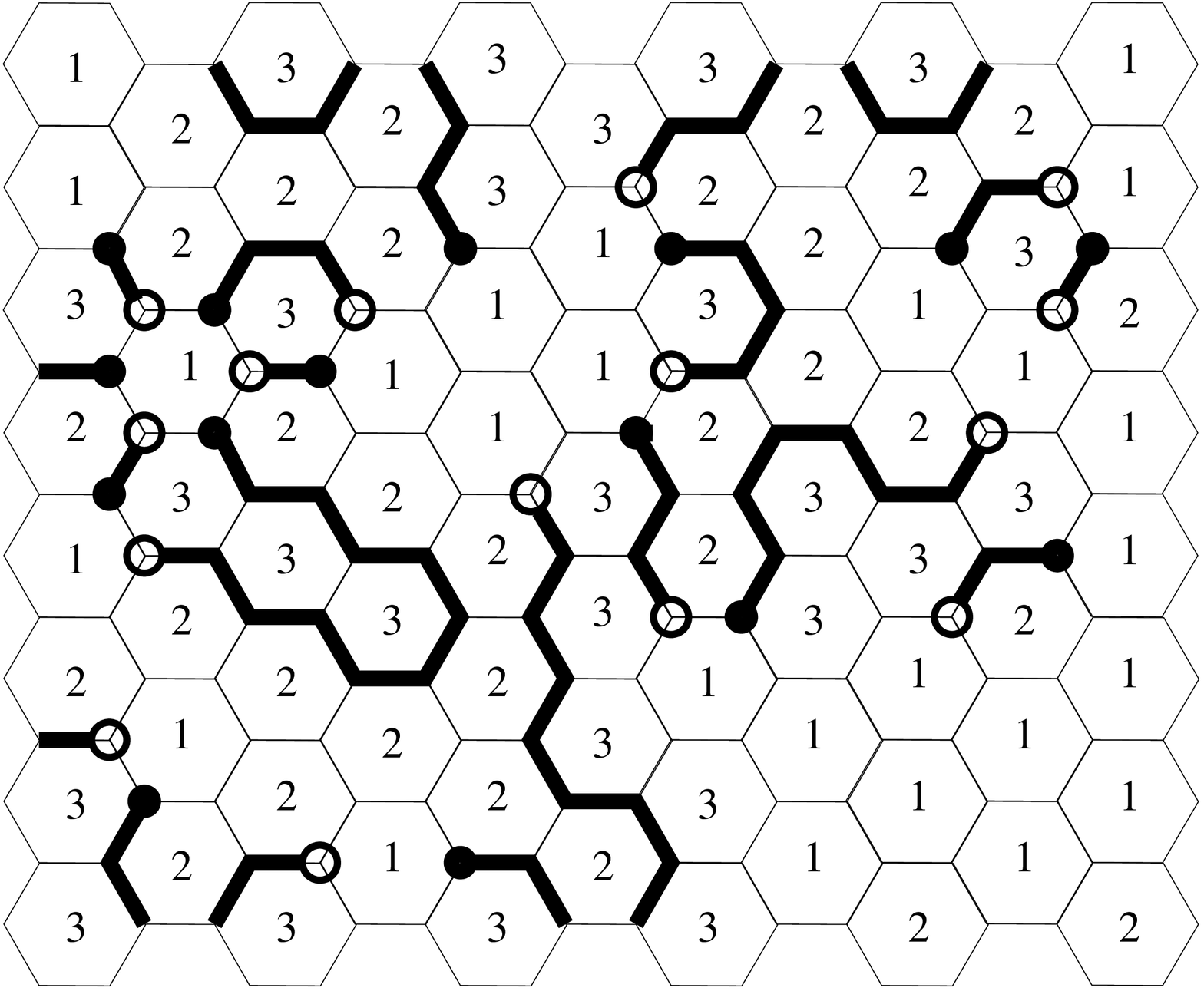}
\\[1mm]
{\bf (b)}
\end{center}
\caption{(a) The circle of axion field expectation values is divided into 
three equal parts with the CP conserving value $\alpha = 0$ in the middle
of part 1. (b) Each cell of a hexagonal grid is randomly assigned 1,2 or 
3. This results in a set of domain walls (thick lines), upgoing strings 
(open circles), and downgoing strings (filled circles), as described in 
the text.}
\begin{center}
\includegraphics[width=13cm]{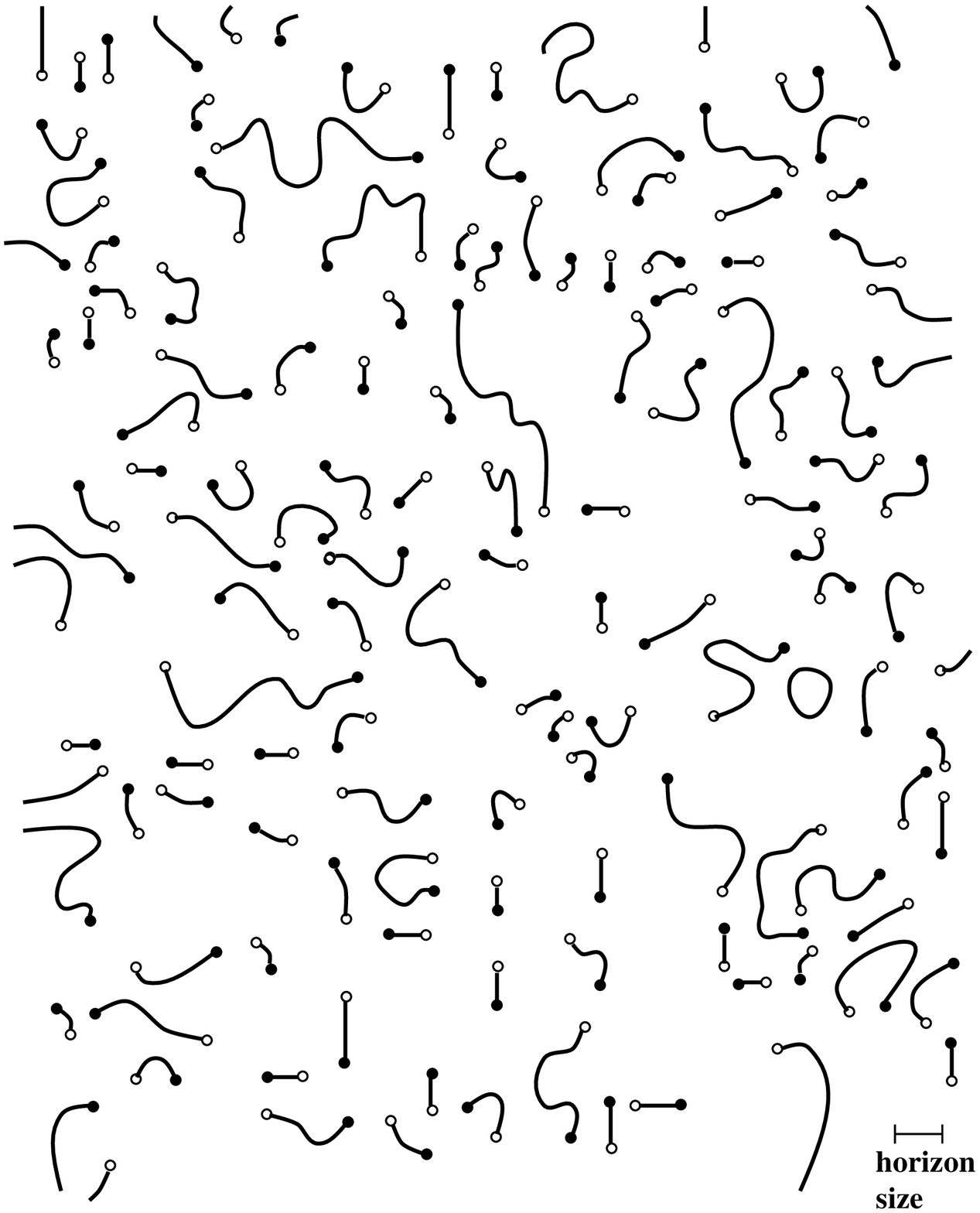}
\end{center}
\caption{Same as in Fig. 2b except that the area covered is larger, the 
underlying grid has been removed and the walls have been smoothed.}
\epsfxsize=130mm
\centerline{\epsfbox{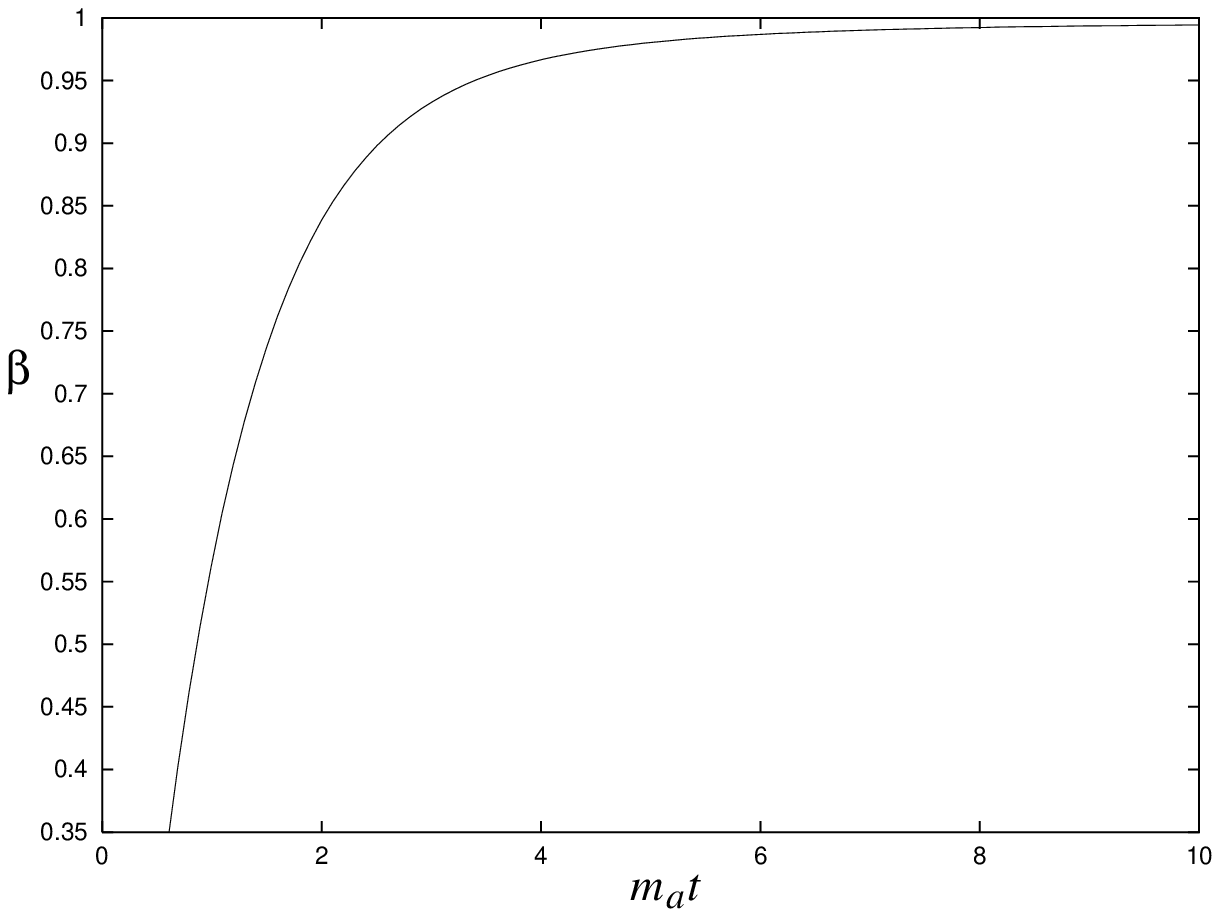}}
\caption{Speed of the string core as a function of time for the case
$1/m_a=400$, $\sqrt{\lambda}/m_a =10$, and $v =0$.}
\label{speed}
\epsfxsize=130mm
\centerline{\epsfbox{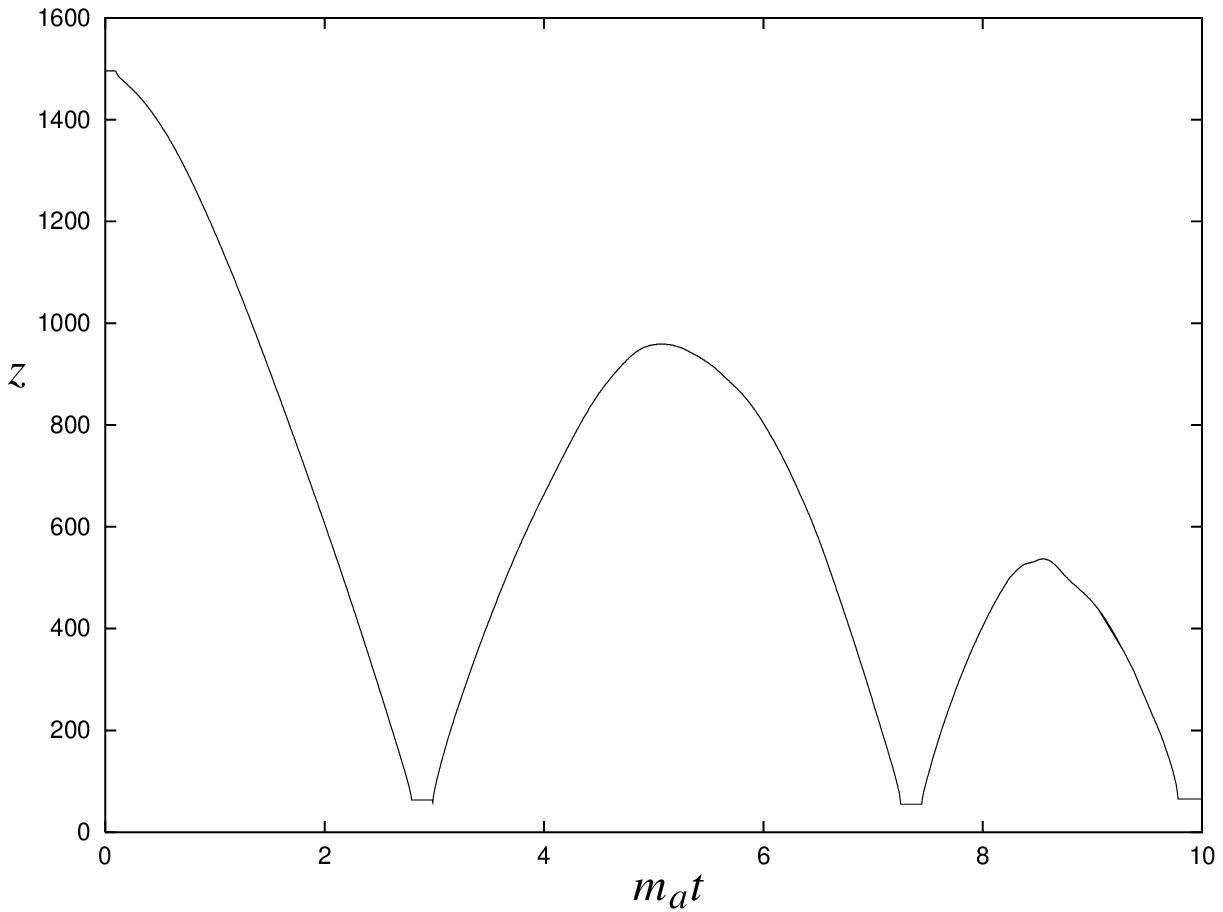}}
\caption{Position of the string or anti-string core as a function of time 
for $1/m_a=1000$, $\lambda = 0.0002, D=2896$, and $v=0$.  The string and
anti-string cores have opposite $z$.  They go through each other and 
oscillate with decreasing amplitude.}
\label{core}
\epsfxsize=130mm
\centerline{\epsfbox{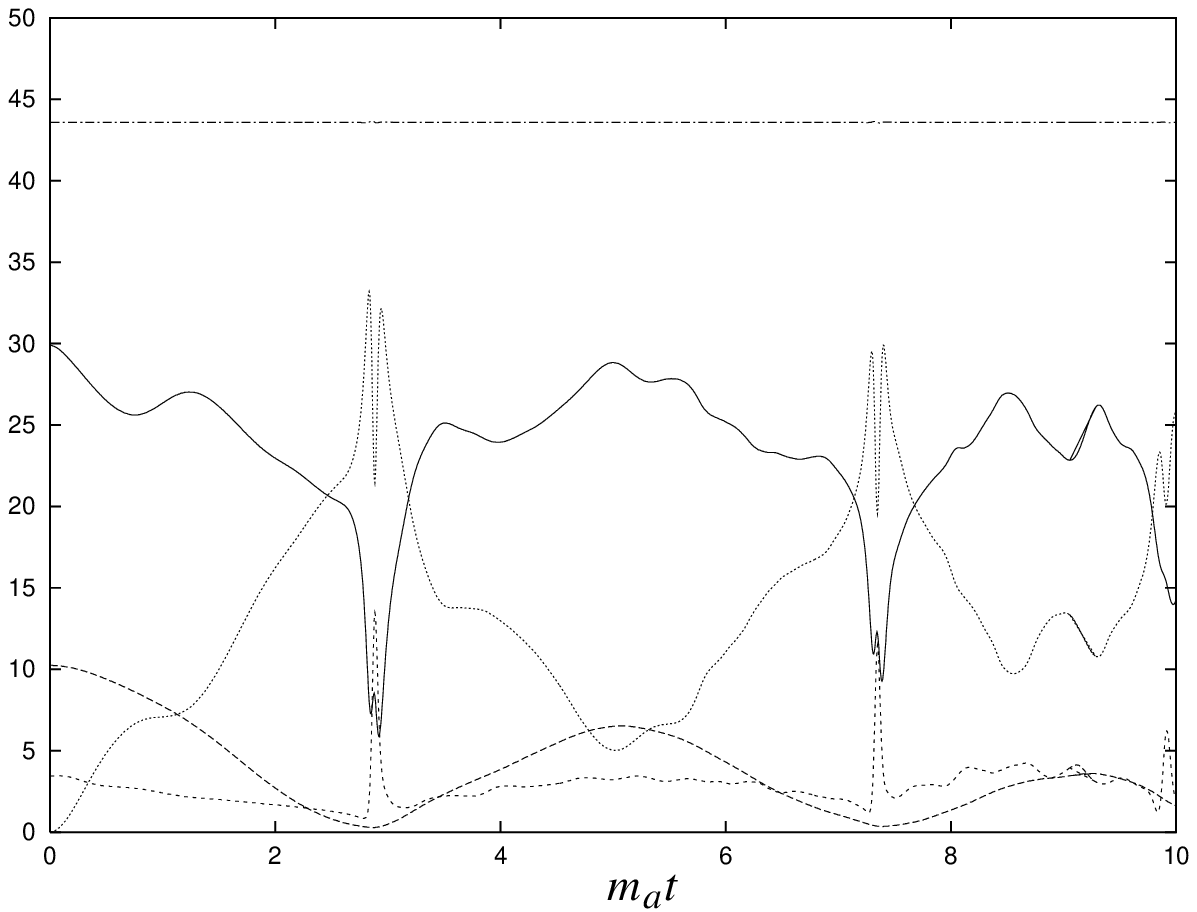}}
\begin{center}
{\bf (a)}
\end{center}
\epsfxsize=130mm
\centerline{\epsfbox{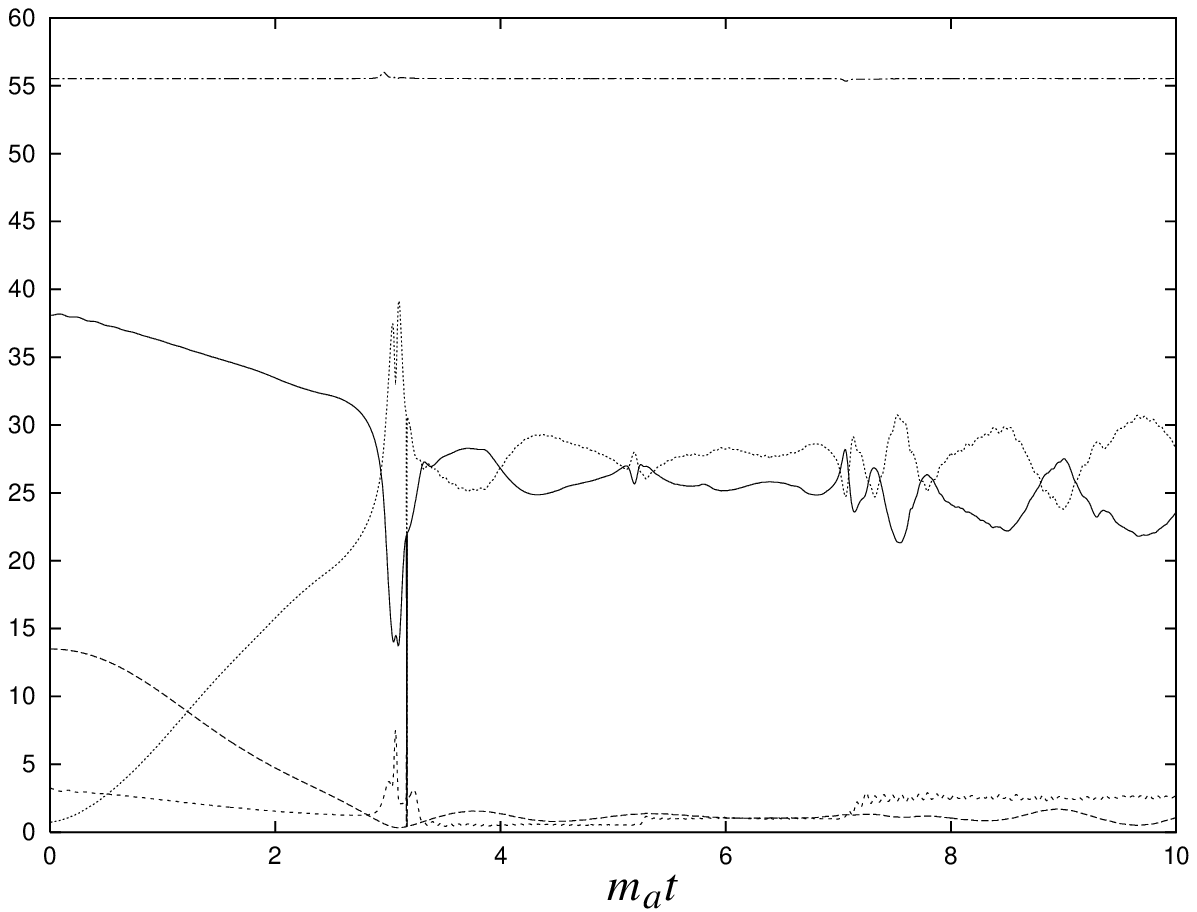}}
\begin{center}
{\bf (b)}
\end{center}
\caption{Gradient (solid), kinetic (dotted), wall potential (long dash),
string potential (short dash) and total (dot dash) energy as a function
of time for the case $1/m_a=1000$, $\lambda = 0.0002$, $D=2896$, and 
$v=0$ (a), and for the case $1/m_a=500$, $\lambda = 0.0032$, $D=2096$ and 
$v=0.25$ (b).}
\epsfxsize=150mm
\centerline{\epsfbox{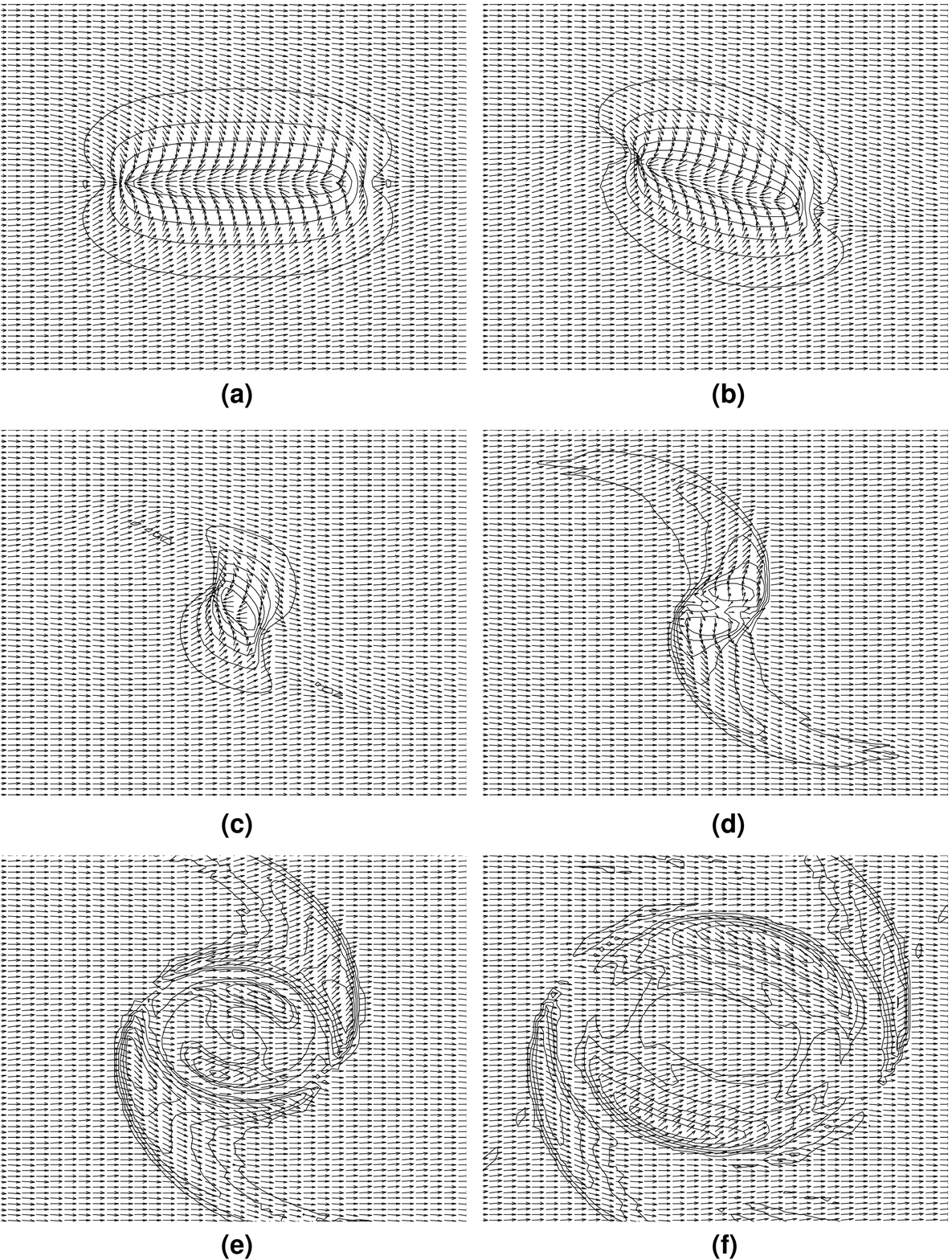}}
\caption{Decay of a wall at successive time intervals $\Delta t = 1.2/m_a$
for the case $m_a^{-1}=100$, $\lambda = 0.01$, $D=524$, and $v = 0.6$.}
\label{rotat}
\epsfxsize=150mm
\centerline{\epsfbox{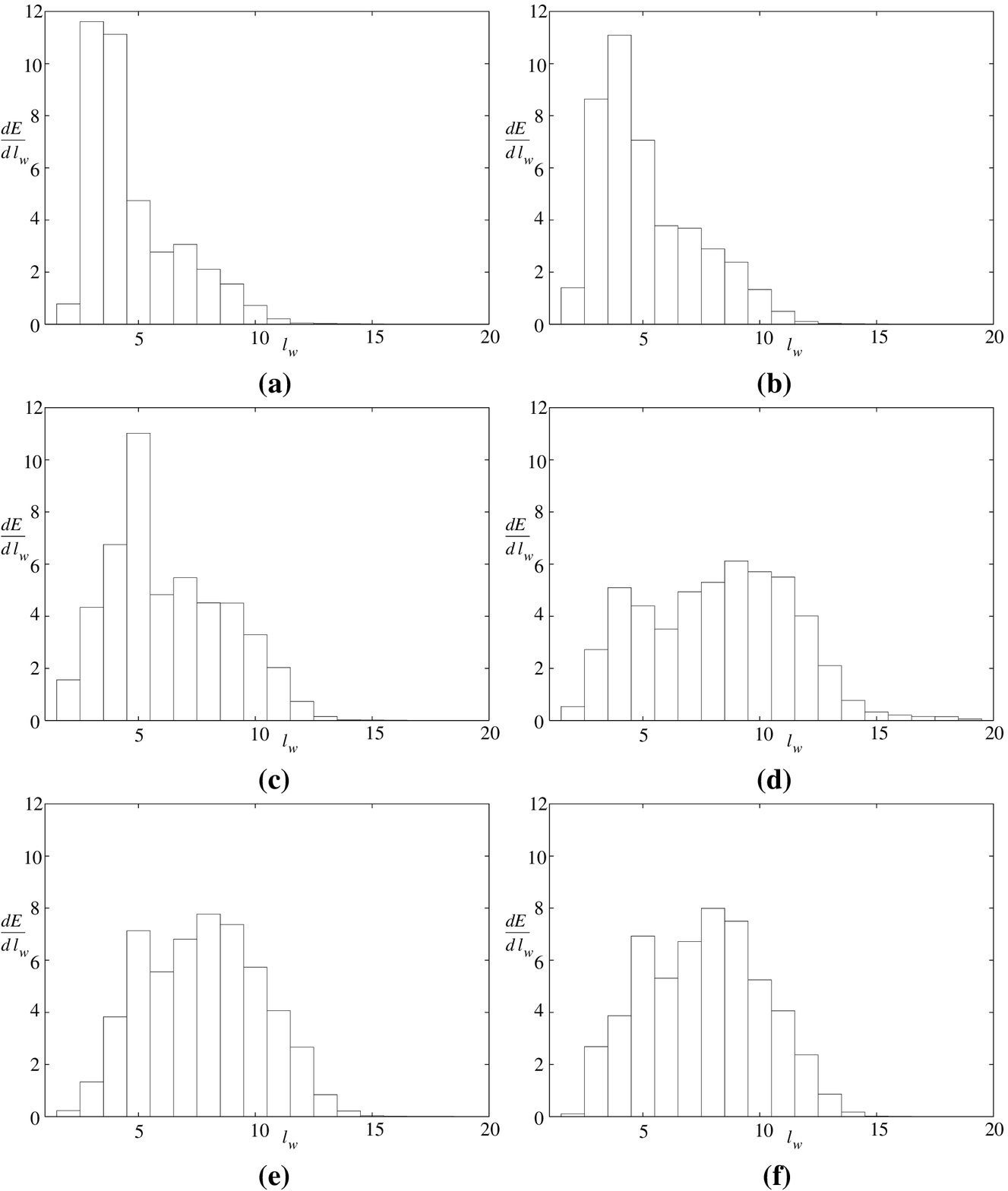}}
\caption{Energy spectrum at successive time intervals $\Delta t = 1/m_a$,
with \\ $l_W=18 \log(w/m_a)/\log(w_{\rm max}/m_a) +2$ and 
$w_{\rm max}= \sqrt{8+m_a^2}$, for the case $m_a^{-1}=500$, $\lambda=0.0032$,
$D=2096$ and $v=0.25$.}
\label{spectrum}

\epsfxsize=135mm
\centerline{\epsfbox{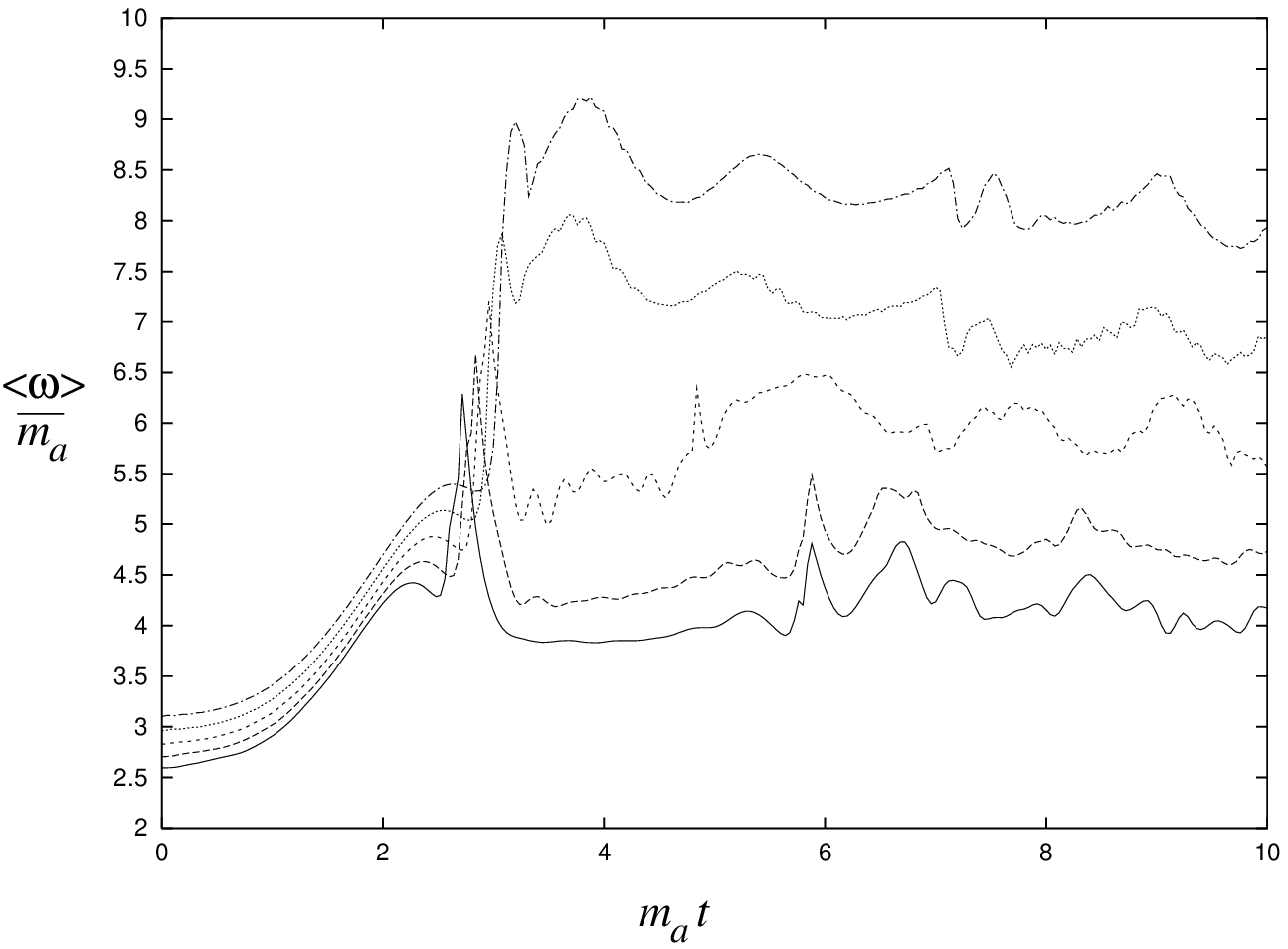}}
\caption{$\langle \omega \rangle$ as a function of time for $1/m_a=500$, 
$D=2096$, $v=0.25$ and $\lambda=$ 0.0004 (solid), 0.0008 (long dash),
0.0016 (short dash), 0.0032 (dot) and 0.0064 (dot dash).  After the wall
has decayed, $\langle \omega \rangle$ is the average energy of radiated
axions.}
\label{energy}

\epsfxsize=135mm
\centerline{\epsfbox{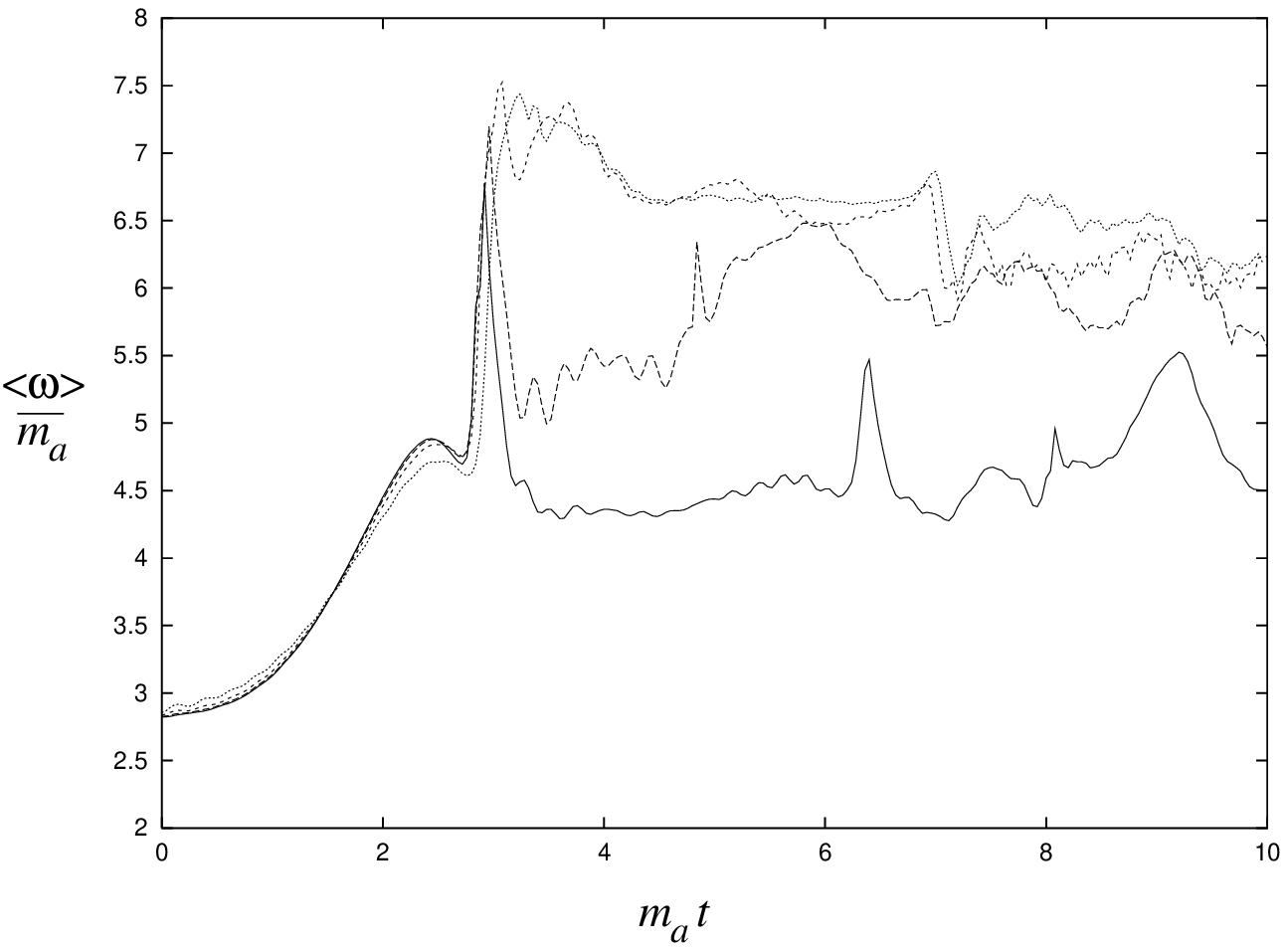}}
\caption{Same as in Fig. 9 except $\lambda = 0.0016$, and $v=$ 0.15 
(solid), 0.25 (long dash), 0.4 (short dash) and 0.6 (dot).}
\end{figure}

\end{document}